\documentclass{article}
\usepackage[utf8]{inputenc}
\usepackage[hidelinks]{hyperref}

\usepackage[english]{babel}
\usepackage[parfill]{parskip} 
\usepackage{float}				
\usepackage{array}
\usepackage{graphicx}
\usepackage{caption}
\usepackage{subcaption}
\usepackage{epstopdf}
\usepackage{comment}
\usepackage{setspace}  
\usepackage[T1]{fontenc}
\DeclareTextFontCommand\textcourier{\fontfamily{qcr}}

\usepackage{booktabs}
\usepackage[table]{xcolor}
\usepackage{bigints}
\usepackage{arydshln}
\usepackage{appendix}
\usepackage{multirow}
\usepackage{amsmath}
\usepackage{mathtools}
\usepackage{subcaption}
\usepackage{bbm}

\usepackage{etoolbox}

\usepackage{lineno} 

\usepackage{appendix}

\usepackage{csquotes}

\usepackage{dsfont}
\usepackage{amsfonts}
\usepackage{amsthm}
\usepackage{amssymb}
\usepackage{lipsum}
\usepackage{geometry}
\usepackage{listings}
\usepackage{helvet}   
\usepackage{booktabs}
\usepackage{natbib}
\PassOptionsToPackage{table,dvipsnames}{xcolor}

\begin{document}


\begin{titlepage}
\begin{center}

    { \huge 
    {\fontfamily{put}\selectfont
    Improved order selection method for hidden Markov models: a case study with movement data\\[0.4cm]} }
        \textsc{\LARGE Fanny Dupont\footnote{Department of Statistics, University of British Columbia, Vancouver, British Columbia V6T 1Z4 Canada}, Marianne Marcoux\footnote{Freshwater Institute, Fisheries and Oceans Canada, 501 University Avenue
Winnipeg, MB R3T 2N6
Canada}, Nigel Hussey\footnote{Department of Integrative Biology, University of Windsor, Windsor, ON N9B 3P4, Canada} and Marie Auger-Méthé\footnote{Department of Statistics, University of British Columbia, Vancouver, British Columbia V6T 1Z4 Canada}\footnote{Institute for the Oceans and Fisheries, University of British Columbia, Vancouver, British Columbia V6T 1Z4 Canada}}\vskip5cm
      \begin{flushleft}
        {Corresponding author: Fanny Dupont\\
        Address: 2207 Main Mall, \\
      Vancouver, BC V6T 1Z4, Canada\\
        Email: fanny.dupont@stat.ubc.ca\\[1cm]}
      \end{flushleft}

 \begin{flushleft}
Running title: Improved order selection in non-stationary HMMs.\\
The authors declare no conflict of interest.
\end{flushleft}
    \vfill
    \end{center}
    \section*{Acknowledgements} 
I acknowledge the support of the Natural Sciences and Engineering Research Council of Canada (NSERC), the Canadian Research Chairs program, BC Knowledge Development fund and Canada Foundation for Innovation’s John R. Evans Leaders Fund, the Canadian Statistical Sciences Institute (CANSSI) as well as the support of Fisheries and Oceans Canada (DFO). I thank the community of Mittimatalik (Pond Inlet) for its support in tagging operations and the devoted people who led operations in the field. Field work was supported by the Polar Continental Shelf Program, Fisheries and Oceans Canada, the Nunavut Wildlife Management Board, the Nunavut Implementation Fund, and World Wildlife Fund Canada. This research was enabled by support provided by Compute Canada (www.computecanada.ca). I am grateful to Dr$.$ Daniel J. McDonald, as well as my committee members Dr$.$ Matías Salibián-Barrera and Dr$.$ Nancy E. Heckman for the constructive suggestions and discussions.

\section*{Author Contribution}
Fanny Dupont and Marie Auger-Methé conceived the ideas and designed the methodology; Marianne Marcoux contributed in ensuring the accurate application of the method in the case study; Nigel Hussey and Marianne Marcoux conducted fieldwork; Fanny Dupont conducted the analyses and prepared the manuscript. MM, MAM, NH contributed to fund the study, and MM and MAM contributed to the supervision. All co-authors provided constructive feedback on written drafts.

Our study engaged with the community of Mittimatalik (Pond Inlet), and involved the Mittimatalik Hunter and Trapper Organization in field and tagging operations. Local individuals played key roles in leading these efforts. While Inuit individuals were involved in data collection for the case study, none were involved in the development of the statistical analysis that is the focus of this paper.

\section*{Data Availability}
All narwhal tracking data used in the study are accessible and available by request through Fisheries and Oceans Canada (Marianne Marcoux, Marianne.Marcoux@dfo-mpo.gc.ca). R code is available from the corresponding author by request.

\end{titlepage}
\doublespacing

\section*{Abstract}
\begin{enumerate}

    \item Hidden Markov models (HMMs) are a versatile statistical framework commonly used in ecology to characterize behavioural patterns from animal movement data. In HMMs, the observed data depend on a finite number of underlying hidden states, generally interpreted as the animal's unobserved behaviour. The number of states is a crucial parameter, controlling the trade-off between ecological interpretability of behaviours (fewer states) and the goodness of fit of the model (more states). Selecting the number of states, commonly referred to as order selection, is notoriously challenging. Common model selection metrics, such as AIC and BIC, often perform poorly in determining the number of states, particularly when models are misspecified.     \vskip0.4cm

    \item  Building on existing methods for HMMs and mixture models, we propose a double penalized likelihood maximum estimate (DPMLE) for the simultaneous estimation of the number of states and parameters of non-stationary HMMs. The DPMLE differs from traditional information criteria by using two penalty functions on the stationary probabilities and state-dependent parameters. For non-stationary HMMs, forward and backward probabilities are used to approximate stationary probabilities.  \vskip0.4cm 
     \item Using a simulation study that includes scenarios with additional complexity in the data, we compare the performance of our method with that of AIC and BIC. We also illustrate how the DPMLE differs from AIC and BIC using narwhal (\textit{Monodon monoceros}) movement data.  \vskip0.4cm 
     
     \item The proposed method outperformed AIC and BIC in identifying the correct number of states under model misspecification. Furthermore, its capacity to handle non-stationary dynamics allowed for more realistic modeling of complex movement data, offering deeper insights into narwhal behaviour. Our method is a powerful tool for order selection in non-stationary HMMs, with potential applications extending beyond the field of ecology.

\end{enumerate}
\vskip0.4cm
\textsc{Keywords\\}
Animal movement, double penalized maximum likelihood estimate (DPMLE), HMM, information criteria, non-stationary, order selection, SCAD\vskip1cm

\textsc{Data Availability\\}
All narwhal tracking data used in the study are accessible and available by request through Fisheries and Oceans Canada (Marianne Marcoux, Marianne.Marcoux@dfo-mpo.gc.ca). R code is available from the corresponding author by request.

\section{Introduction}
\label{Introduction}

Understanding animals' movement and behaviour is crucial for conservation, and is therefore driving the rapid development of animal movement modelling (\citealp{sutherland_importance_1998}).  One of the most popular modelling approaches is the Hidden Markov Model (HMM), a versatile statistical tool for modelling time series (\citealp{patterson_using_2010}; \citealp{mcclintock_uncovering_2020}). Advances in tracking technology and biologging data resolution have increased HMMs' popularity in movement ecology and they are now widely used to infer animals' behaviour from telemetry data (\citealp{zucchini_hidden_2017}; \citealp{glennie_hidden_2023}). HMMs assume that the observed time series arises from a sequence of unobserved states, evolving in a finite state space. The states usually carry information about the phenomenon of interest, such as the survival status (\citealp{mcclintock_uncovering_2020}) or the behavioural state of an animal (\citealp{morales_extracting_2004}). Movement ecologists commonly seek to study the environmental conditions that may trigger switches between these states (e.g., time of the day in \citealp{ngo_understanding_2019}). 

In practice, when fitting an HMM to movement data, the parameters for both the observation and the hidden processes, along with the number of hidden states, need to be estimated. While HMMs can be applied to data using various inferential frameworks (see \citealp{leos-barajas_introduction_2018}; \citealp{augermethe_guide_2021} for Bayesian methods), maximum likelihood inference is the most commonly used approach to fit HMMs in ecology (\citealp{zucchini_hidden_2017}; \citealp{mcclintock_worth_2021}) and, as such, is the focus of this paper. However, selecting the number of states, known as order selection, remains notoriously challenging (\citealp{pohle_selecting_2017}). Order selection is crucial to control the trade-off between interpretability and the goodness of fit of the model. The lack of universal tools often leads researchers to rely on model validation techniques or their own expertise to choose, sometimes arbitrarily, an appropriate number of states. Consequently, many practitioners choose a number of states that is easily interpretable (\citealp{deruiter_multivariate_2017}). This order selection problem with HMMs is similar to the challenges associated with estimating the number of components of finite mixture models (\citealp{chen_order_2008}).

 When estimating the parameters by maximum likelihood, the standard tools to choose between models with different numbers of states are the likelihood ratio test and information criteria (e.g., Akaike Information Criterion – AIC, Bayesian Information Criterion – BIC). Existing likelihood ratio methods cannot effectively test for order selection in HMMs, leading many researchers to rely on information criteria (\citealp{dannemann_testing_2008}).
  
 
The conventional approach for order selection in HMMs involves a two-stage procedure: fitting models with varying numbers of states and selecting the best-fitting model with model selection criteria (\citealp{celeux_selecting_2008}). AIC and BIC often perform poorly in selecting the order of HMMs, particularly when applied to misspecified models (i.e., the true generating process in not among the models compared; \citealp{pohle_selecting_2017}; \citealp{li_incorporating_2017}). Alternative model selection criteria have been proposed. The integrated completed likelihood criterion (ICL; \citealp{biernacki_assessing_2000}) is promising, but is too sensitive to the overlap between the state-dependent distributions and tends to underestimate the order (\citealp{pohle_selecting_2017}). The cross-validated likelihood criterion circumvents the theoretical challenges associated with information criteria (\citealp{smyth_model_2000}), but is computationally challenging and does not outperform BIC (\citealp{celeux_selecting_2008}). 

Two-stage procedures can be computationally expensive, particularly as model complexity and sample size increase. Such a situation is likely to arise with increasingly complex animal movement data. \cite{de_chaumaray_estimation_2022} introduced a one-stage approach for order selection in non-parametric HMMs, avoiding the inefficiencies and computational burden of a two-stage procedure. However, it cannot accommodate models with time-varying covariates (hence non-stationary HMMs). Penalized maximum likelihood estimators are alternative one-stage methods (\citealp{mackay_estimating_2002}; \citealp{gassiat_optimal_2003}). \cite{chen_order_2008} argued that while some penalized methods for order selection in HMMs prevent \textit{overfitting of type I} (i.e., estimating states that animals rarely occupy), they do not address \textit{overfitting of type II}, where component densities overlap. They proposed a method using the smoothly clipped absolute deviation (SCAD; \citealp{fan_variable_2001}) function to overcome both overfitting types in mixture models. However, no methods exist for non-stationary models, highlighting the need for a universal statistical method for order selection in non-stationary HMMs.

This paper introduces an intuitive double penalized maximum likelihood estimate (DPMLE) for the simultaneous estimation of the order and parameters of non-stationary HMMs. It is built on the method proposed in \cite{chen_order_2008} to estimate the order of mixture models. \cite{hung_hidden_2013} introduced it for stationary HMMs, where it outperformed AIC and BIC in selecting the right number of states. We go further by extending this framework to non-stationary HMMs, which incorporate time-varying covariates in the hidden process, and evaluate its performance using a simulation study with misspecification, since these situations are common with animal movement data (\citealp{pohle_selecting_2017}; \citealp{li_incorporating_2017}). This is the first time that a double penalized likelihood method for order selection in non-stationary HMMs is explored, and evaluated under model misspecification.

We will illustrate the use of the DPMLE using movement data from narwhal (\textit{Monodon monoceros}),  a species vulnerable to climate change, specifically by decreasing sea ice and associated increasing predator presence and ship traffic (\citealp{pizzolato_changing_2014}; \citealp{breed_sustained_2017}). Previous studies identified "distance to shore" as an important covariate for narwhal habitat selection (\citealp{kenyon_baffin_2018}; \citealp{ngo_understanding_2019}; \citealp{shuert_divergent_2023}). Thus, we investigate non-stationary HMMs that include distance to shore, and compare the results found when selecting the number of states with our new DPMLE method and AIC and BIC. To our knowledge, the application of a double penalized likelihood method has not yet been explored in animal movement analyses.

\section{Materials and Methods}
\label{MaterialsMethods}
\subsection*{Hidden Markov Models}
\label{HMM}


Consider a $N^{\star}$-state HMM describing $M$ independent individuals, where $\boldsymbol{Y}=(\boldsymbol{Y}^{1}_{1:T_1},\ldots,\boldsymbol{Y}^{M}_{1:T_M}),$ with $\boldsymbol{Y}^m_{1:T_m} =(\boldsymbol{Y}^m_1,\ldots,\boldsymbol{Y}^m_{T_m})$ the observations of individual $m$
and $\boldsymbol{S}^{m}_{1:T_m}=(S_1^m,\ldots,S^m_{T_m})$ the associated sequence of hidden states, $S_t^{m} \in \{1,2,\ldots,N^{\star}\}.$ Transitions between hidden states are driven by the $N^{\star}\times N^{\star}$ transition probability matrix (tpm) $\boldsymbol{\Gamma}=(\gamma_{ij})_{i,j\in [1,N^{\star}]\times [1,N^{\star}]}:$
\begin{center}
    \begin{equation}
    \label{constraint1}
      \mathbb{P}({S}^m_{t+1} = i|{S}^m_t = j)=\gamma_{ij}, \qquad \underset{j=1}{\overset{N^{\star}}{\sum}}\gamma_{ij}=1,
    \end{equation}
\end{center}
and satisfy the Markov property. The initial state distribution, $\boldsymbol{\delta}$, satisfies 
\begin{equation}\label{constraint2}\underset{i=1}{\overset{N^{\star}}{\sum}}\delta_i=1.\end{equation} 

$\boldsymbol{S}^m_{1:T_m}$ is homogeneous (i.e., time-independent). If there exists a row vector $\boldsymbol{\pi}$ that satisfies $\boldsymbol{\pi}\boldsymbol{\Gamma}=\boldsymbol{\pi}$, the chain is further said to be \textit{stationary} and we commonly set $\boldsymbol{\delta}=\boldsymbol{\pi}$ (\citealp{zucchini_hidden_2017}). The stationary probability, $\boldsymbol{\pi}$, can be interpreted as the proportion of time an individual spends in each state and is entirely determined by the tpm (\citealp{zucchini_hidden_2017}). If $S^m_t = i$,  the conditional density of $\boldsymbol{Y}^m_t$ is $f_i(\cdot;\boldsymbol{\theta}_i)$, where $\boldsymbol{\theta}_i$ is a state-dependent parameter describing the emission (also called \textit{state-dependent})
distribution. The observations are assumed independent given the states.


The likelihood can be written as follows (\citealp{zucchini_hidden_2017}):\\
\begin{center}
\begin{equation}
\label{nullmulti}
   \mathcal{L}_M(\boldsymbol{\Psi}|\boldsymbol{y}) \coloneqq \mathcal{L}_M = \underset{m=1}{\overset{M}{\prod}}\boldsymbol{\delta} \boldsymbol{\Gamma} \textbf{P}(\boldsymbol{y}^{m}_{1})\boldsymbol{\Gamma} \textbf{P}(\boldsymbol{y}^m_{2})\ldots\boldsymbol{\Gamma} \textbf{P}(\boldsymbol{y}^m_{T_m})\textbf{1},  
\end{equation}
\end{center}
with vector of model parameters $\boldsymbol{\Psi}=(\boldsymbol{\delta},\boldsymbol{\Gamma},\Theta),$ ${\Theta}= (\boldsymbol{\theta}_{1},\ldots,\boldsymbol{\theta}_{N^{\star}})$ and $\textbf{P}(\boldsymbol{y}^m_{t})$ a $N^{\star} \times N^{\star}$ diagonal matrix with $(i,i)^{th}$ entry $f_i(\boldsymbol{y}^m_{t};\boldsymbol{\theta}_{i})$.  

 We consider a frequentist approach, which involves obtaining maximum likelihood estimates $\widehat{\boldsymbol{\Psi}}= (\hat{\boldsymbol{\delta}},\hat{\boldsymbol{\Gamma}},\hat{{\Theta}})={\text{argmax}}_{\boldsymbol{\Psi}}{ \mathcal{L}_M(\boldsymbol{\Psi}|\boldsymbol{y})}$. We use the Expectation-Maximization (EM) algorithm (\citealp{baum_maximization_1970}). An EM iteration consists of an expectation (E) step, and a maximization (M) step. It is based on the rewriting the likelihood in terms of forward probabilities ${\alpha}_t^m(i) = \mathbb{P}(\boldsymbol{Y}^m_{1:t} = \boldsymbol{y}^m_{1:t}  ; S^m_t = j)$ and backward probabilities ${\beta}_t^m(i)=\mathbb{P}(\boldsymbol{Y}^m_{t+1:T_m} = \boldsymbol{y}^m_{t+1:T_m}  ; S^m_t = j)$, for all $m \leq M, t\leq T_m, i \leq N^{\star}$. 
 By combining forward and backward probabilities, we can efficiently compute the posterior probability of being in a specific state at a specific time, given the entire observed sequence. Thus Eq (\ref{nullmulti}) can be rewritten as follows (\citealp{zucchini_hidden_2017}):
\begin{equation}
  \mathcal{L}_M= \underset{m=1}{\overset{M}{\prod}}\underset{i=1}{\overset{N^{\star}}{\sum}}{\alpha}_t^m(i){\beta}_t^m(i).
\end{equation}
Both $\boldsymbol{\alpha}^m_t$ and $\boldsymbol{\beta}^m_t$ can be computed recursively using the forward-backward algorithm as follows:
\begin{center}
$
\begin{array}{lll}
{\alpha}_t^m(j) = \underset{i = 1}{\overset{N^{\star}}{\sum}}{\alpha}_{t-1}^m(i)\Gamma_{i,j}f_j(y_t^m;\boldsymbol{\theta}_j), \qquad t = 2,\ldots T_m, m \leq M,  &\text{ and }& {\alpha}_1^m(j) = \delta_i f_j(y_1^m;\boldsymbol{\theta}_j).\\
{\beta}_t^m(j) = \underset{i = 1}{\overset{N^{\star}}{\sum}}\Gamma_{j,i}f_i(y_{t+1}^m;\boldsymbol{\theta}_j){\beta}_{t+1}^m(i),  \qquad t = 1,\ldots T_m-1, m \leq M,  &\text{ and }& {\beta}_{T_m}^m(i) = 1.
\end{array}
$\end{center}
\vskip0.5cm

The model described by Eq (\ref{nullmulti}) is a standard HMM, that can be further extended to incorporate individual heterogeneity and covariates in both processes. 
In animal movement, covariates are generally included in the hidden process
(non-homogeneous HMM; \citealp{mckellar_using_2015}; \citealp{leos-barajas_introduction_2018}), via a multinomial logit link as follows:
\begin{center}
\begin{equation}
\label{t.p.m.t}
    \gamma_{ij}^{(t)} = \frac{e^{c_{ij}^{(t)} }}{\underset{k}{\sum}{e^{c_{ik}^{(t)}}}},
\end{equation}
\text{with}
\end{center}

\begin{center}
\begin{equation}  
\label{t.p.m.tt}
c_{ij}^{(t)}= 
\left\{
\begin{array}{cc}
\beta_0^{ij}+\underset{c=1}{\overset{C}{\sum}}\beta_c^{ij}\omega_c^{(t)},& \text{ for } i\neq j.  \\

   0 & \text{ otherwise,} 
\end{array}
\right.
\end{equation}
\end{center}
where $(\omega_1^{(t)},\ldots,\omega_C^{(t)})$ is the vector of the $C$ covariates at time $t$ and $\boldsymbol{\beta}^{ij}=({\beta}^{ij}_0,\ldots,{\beta}^{ij}_C)$ is the vector of regression coefficients for the transition probability $\gamma_{ij}^{(t)}$, $t\geq0$. The vector of parameters to estimate becomes $\boldsymbol{\Psi}= (\boldsymbol{\delta},\boldsymbol{\beta},\boldsymbol{\mu},\boldsymbol{\sigma})$. Non-homogeneous HMMs are, by definition, non-stationary.


Key features of interest in animal movement modelling include the state-dependent distributions, which describe movement patterns (e.g., speed, tortuosity), and the temporal structure, reflected in the transition probability matrix (e.g., behavioural persistence from diagonal entries). Time-varying covariates help explain movement patterns and environmental drivers of behavioural transitions (\citealp{mckellar_using_2015}; \citealp{florko_review_nodate}). 
\vskip0.4cm

{\textsc{Model selection criteria}}

Model selection is a crucial step to check for prediction accuracy and understand mechanisms driving behaviour. We focus on the AIC (\citealp{akaike_new_1974}) and BIC (\citealp{schwarz_estimating_1978}) since they are commonly used by ecologists (\citealp{augermethe_guide_2021}). 
Although they arise from different approaches, they have similar formulae. AIC is defined as follows:
\begin{center}
\begin{equation}
\label{AIC}
    \text{AIC} = -2\log { \mathcal{L}(\hat{\boldsymbol{\Psi}}|\boldsymbol{y})}+2k,
\end{equation}
\end{center}
where $k$ is the number of estimated parameters in the model, $\boldsymbol{y} $ the observed data and $\widehat{\boldsymbol{\Psi}}= {\text{argmax}}_{\boldsymbol{\Psi}}{ \mathcal{L}(\boldsymbol{\Psi}|\boldsymbol{y})}$ the MLE. BIC can be computed as follows: 
\begin{center}
\begin{equation}
\label{bic}
    \text{BIC} = -2\log { \mathcal{L}(\hat{\boldsymbol{\Psi}}|\boldsymbol{y})}+k\log(n),
\end{equation}
\end{center}
where $n$ is the number of observations. Lower values for AIC and BIC indicate a better model. Both criteria penalize the number of parameters through the second term of the equations. Since its penalty function increases with the sample size, BIC tends to favour more parsimonious models than AIC. This behaviour arises in our simulation study (see Tables in supporting information).\vskip0.4cm

Theoretical justifications for the use of AIC and BIC for order selection in HMMs have not been derived yet and one should be cautious when using them. States in HMMs can be viewed as random effects, leading to challenges such as boundary problems (\citealp{bolker_generalized_2009}). HMMs also fail to satisfy the standard assumptions underlying the derivation of both information criteria, making them unreliable in selecting the best HMM. For example, under overestimation ($\widehat{N}>N^{\star}=2$), the assumption of a unique MLE does not hold (\citealp{watanabe_widely_2013}; \citealp{drton_bayesian_2016}). In addition to these theoretical problems, AIC and BIC perform poorly under model misspecification. They tend to select additional states to compensate for the structure missing due to model misspecification (\citealp{pohle_selecting_2017}; \citealp{li_incorporating_2017}). 

\vskip0.4cm

{\subsection*{Likelihood-based double penalized method for order selection}}
\label{penalizedmethode}

 We propose a double penalized maximum likelihood estimate (DPMLE) of the number of hidden states, based on the method developed by \cite{chen_order_2008} and \cite{hung_hidden_2013}, adapted to non-stationary HMMs. The main idea of the DPMLE is to remove virtually empty states (\textit{overfitting of type I}) and merge duplicate states together (\textit{overfitting of type II}).\vskip0.4cm

Consider a stationary $N^{\star}$-state HMM described by Eq (\ref{nullmulti}). The true number of states $N^{\star} $ is unknown and must be estimated. From the order's upper bound $N > N^{\star}$, the double penalized method estimates, by minimizing two penalty functions, a lower or equal order by first clustering and then merging similar states 
together.

Without loss of generality, we describe the double penalized procedure with state-dependent distributions characterized by two parameters: $\boldsymbol{\theta}_i = (\mu_i,{\sigma}_i), i \leq N$. The double penalized log-likelihood function is defined as:

\begin{center}
\begin{equation}
\label{doublepenalized}
\tilde{l}_M(\boldsymbol{\Psi}|\boldsymbol{y})= l_M(\boldsymbol{\Psi}|\boldsymbol{y}) + C_{N}\underset{j=1}{\overset{N}{\sum}}\log{\pi_j}- \underset{j=1}{\overset{N-1}{\sum}}p_{\lambda_M}({\eta_j}),
\end{equation}
\end{center}
with $\eta_j = \mu_{j+1}-\mu_{j}, $ for $j > 1$, $\mu_1\leq \mu_2 \leq \ldots \leq \mu_{N}$, $C_N$ is a constant $> 0$, $l_M(\boldsymbol{\Psi}|\boldsymbol{y}) = \log{\mathcal{L}_M(\boldsymbol{\Psi}|\boldsymbol{y})}$ and
$p_{\lambda_M}$ is a penalty function.

The double penalized maximum likelihood method penalizes (1) the stationary probabilities to prevent small values of stationary probabilities (\textit{overfitting of type I}) and (2) the state-dependent parameters to penalize the overlap between the component distributions (\textit{overfitting of type II}). The DPMLE is then:

\begin{equation}
\begin{aligned}
\widehat{\boldsymbol{\Psi}}_{\text{DPMLE}} &= \underset{\boldsymbol{\Psi}}{\text{argmax}} \, \left( l_M(\boldsymbol{\Psi}|\boldsymbol{y}) + C_{N} \sum_{j=1}^{N} \log \pi_j - \sum_{j=1}^{N-1} p_{\lambda_M}(\eta_j) \right), \\
\text{with} \quad \widehat{N} &= \text{number of distinct values of } \{\hat{\mu}_{1},\ldots,\hat{\mu}_{N}\}.
\end{aligned}
\end{equation}

 The penalty term on the left is used to prevent the $\pi_j$'s (i.e., stationary probabilities) with small values, thus penalizes states in which the process spends little time. The penalty term on the right uses a non-negative function $p_{\lambda_M}$ that shrinks small $\eta_k$'s to $0$, thus preventing overfitting resulting from having multiple similar states (i.e., overlapping state-dependent distributions). In animal movement analyses, $p_{\lambda_M}$ could be applied to the mean parameter, as \textit{overfitting of type II} generally concerns emission distributions with close means, rather than overlapping tails (see \cite{hung_hidden_2013} for an example of its use on $\boldsymbol{\sigma}$ on Gaussian HMMs).
 
The DPMLE is consistent in estimating the order and parameters of stationary HMMs with penalty functions $p_{\lambda_M}$ constant outside a small neighbourhood of $0$. Thus, we use the SCAD function and refer to \cite{hung_hidden_2013} for consistency properties of the order and parameter estimates in stationary HMMs.  
The SCAD function is usually characterized by its derivatives as follows:
\begin{center}
\begin{equation}
    p_{\lambda_M}^{\prime}(\eta_j)=M\lambda_M\mathbbm{1}_{\eta_j\leq\lambda_M }+M\lambda_M\frac{(a\lambda_M-\eta_j)_{+}}{(a-1)\lambda_M}\mathbbm{1}_{\eta_j>\lambda_M},
\end{equation}
\end{center}
where $\mathbbm{1}_{(\cdot)}$ is the characteristic function, $a$ a constant $> 2$ and $\lambda_M$ a hyperparameter.

For the remainder of this work, we assume the regularity conditions necessary for the consistency of the stationary DPMLE are met although they are often violated in practice (\citealp{chen_consistency_2005}). 

\vskip0.4cm

{\textsc{Estimation}}

\textit{Stationary Markov chain}\label{EMestimation}

Conditionally on $\lambda_M$ and $C_N$, the algorithm adapts the EM procedure for stationary HMMs. It follows the method of \cite{hung_hidden_2013}, with a modified update for stationary probabilities, which are estimated from the tpm (\citealp{zucchini_hidden_2017}). Consider the framework defined in Eq (\ref{doublepenalized}) with $M=1$. We drop the individual index for convenience, and the double penalized log-likelihood function to maximize is:
    $\tilde{l}(\boldsymbol{\Psi}|\boldsymbol{y})= l(\boldsymbol{\Psi}|\boldsymbol{y}) + C_{N}\underset{j=1}{\overset{N}{\sum}}\log{\pi_j}- \underset{j=1}{\overset{N-1}{\sum}}p_{\lambda}({\eta_j}).$ The complete data double penalized log-likelihood can be written as follows:
\begin{center}
\begin{equation}
  \begin{multlined}
\label{EMeq} 
    \underset{j=1}{\overset{N}{\sum}}u_{j}(1)\log{\pi_{j}}
    + C_N\underset{j=1}{\overset{N}{\sum}}\log{\pi_{j}}
    +  \underset{t=2}{\overset{T}{\sum}}
    \underset{j,i=1}{\overset{N}{\sum}}v_{ij}(t)\log{\gamma_{ij}}
     \\
        +    \underset{t=1}{\overset{T}{\sum}} \underset{j=1}{\overset{N}{\sum}}u_{j}(t)\log{f_j(\boldsymbol{y}_{t};\boldsymbol{\theta}_j)}
    - \underset{j=1}{\overset{N -1}{\sum}}p_{\lambda}({\eta_j}). \quad
    \end{multlined}
\end{equation}
\end{center}
with $u_{j}{(t)}=1 $ if and only if $s_{t}=j,$ for $t \leq T
$ and $v_{jk}(t)=1$ if and only if $s_{t}=j$ and $s_{t-1}=k$, for $t \in (2,\ldots,T)$. 
The E-Step at iteration $p+1$ replaces $u_{i}(t)$ and $v_{ij}(t)$ by their conditional expectations based on the previous parameter estimates $\widehat{\boldsymbol{\Psi}}^{(p)}$ and the observations, allowing us to perform the maximization step as if the states were known. We derive: 
\begin{equation}
\label{UEM}
\left\{
\begin{aligned}
\hat{u}_{j}^{(p+1)}(t) &= \mathbb{E}[u_{j}(t)|\widehat{\boldsymbol{\Psi}}^{(p)},\boldsymbol{Y}=\boldsymbol{y}] = \frac{\hat{\alpha}_{t}^{(p)}(j)\hat{\beta}_{t}^{(p)}(j)}{\mathcal{L}(\hat{\boldsymbol{\Psi}}^{(p)}|\boldsymbol{y})}, \\
\hat{v}_{jk}^{(p+1)}(t) &= \mathbb{E}[v_{jk}(t)|\widehat{\boldsymbol{\Psi}}^{(p)},\boldsymbol{Y}=\boldsymbol{y}] = \hat{\alpha}_{t-1}^{(p)}(j)\hat{\gamma}^{(p)}_{jk}f_k(\boldsymbol{y}_{t};\hat{\boldsymbol{\theta}}_k^{(p)})\hat{\beta}_{t}^{(p)}(k).
\end{aligned}
\right.
\end{equation}
 The notation $\hat{\boldsymbol{\beta}}_{t}^{(p)}$ and $\hat{\boldsymbol{\alpha}}_{t}^{(p)}$ indicates that the parameter estimates from the $p^{th}$ iteration were used for the derivation. Eq (\ref{EMeq}) at iteration $p+1$ becomes:

\begin{center}
\begin{equation}
  \begin{multlined}
\label{EMeq2} 
    \underbrace{\underset{j=1}{\overset{N}{\sum}}\hat{u}_{j}^{(p+1)}(1)\log{\pi_{j}}
    + C_N\underset{j=1}{\overset{N}{\sum}}\log{\pi_{j }}
    +  \underset{t=2}{\overset{T}{\sum}}
    \underset{j,i=1}{\overset{N}{\sum}}\hat{v}_{ij}^{(p+1)}(t)\log{\gamma_{ij}}}_{\text{(I)}}
     \\
        +    \underbrace{\underset{t=1}{\overset{T}{\sum}} \underset{j=1}{\overset{N}{\sum}}\hat{u}_{j}^{(p+1)}(t)\log{f_j(\boldsymbol{y}_{t};\boldsymbol{\theta}_j)}
    - \underset{j=1}{\overset{N-1}{\sum}}p_{\lambda}({\eta_j})}_{\text{(II)}} \quad
    \end{multlined}
\end{equation}
\end{center}

\vskip0.3cm

We start by maximizing (I) with respect to $\boldsymbol{\Gamma}$ under the constraint $\underset{j=1}{\overset{N}\sum}\pi_j=1$. We improve the algorithm proposed by \cite{hung_hidden_2013} by taking into account the fact that estimating $\boldsymbol{\pi}$ essentially consists of estimating $\boldsymbol{\Gamma}$ and derive $\boldsymbol{\pi}$ as follows: 
\begin{center}
\begin{equation}
   \boldsymbol{\pi}=\boldsymbol{1} (I_{N}-\boldsymbol{\Gamma}+\boldsymbol{U})^{-1}, 
\end{equation}
\end{center} with $\boldsymbol{U}$ the $N\times N$ matrix of ones.  

Any non-penalized parameter (e.g., $\boldsymbol{\sigma}$) uses the classic EM algorithm, which usually requires numerical methods. We then maximize (II) with respect to ${\Theta}$. 
Let $\hat{\boldsymbol{\Psi}}^{(p+\frac12)}=(\hat{\boldsymbol{\pi}}^{(p+1)},\hat{\boldsymbol{\Gamma}}^{(p+1)},\hat{\boldsymbol{\mu}}^{(p)},\hat{\boldsymbol{\sigma}}^{(p+1)})$ be the vector of all updated parameters up to this point, at iteration $p+1$. 
To obtain $\hat{\boldsymbol{\mu}}^{(p+1)}=(\hat{\mu}_1^{(p+1)},\ldots,\hat{\mu}_N^{(p+1)} )$, we maximize a local approximation of the SCAD penalty evaluated in $\hat{\boldsymbol{\Psi}}^{(p+\frac12)}$ with respect to $\boldsymbol{\mu}$. This circumvents the non-smoothness of the SCAD penalty (\citealp{zou_one-step_2008}). 
The maximization problem can then be written as follows:
\begin{center}
\begin{equation}
\label{Mstepstatedep}
  {\text{argmax}}_{\boldsymbol{\mu}}\underset{t=2}{\overset{T}{\sum}} \underset{j=1}{\overset{N}{\sum}}\hat{u}_{j}^{(p+\frac12)}(t)\log{f_j(\boldsymbol{y}_{t};\mu_j, \hat{\sigma}_j^{(p+1)})}
    - \underset{j =1}{\overset{N-1}{\sum}}[\underbrace{p_{\lambda}({\hat{\eta}_j^{(p)}})+p_{\lambda}^{\prime}({\hat{\eta}_j^{(p)}})(\eta_j-\hat{\eta}_j^{(p)})}_{\tilde{p}_{\lambda}({\eta}_j;\hat{\eta}_j^{(p)})}].  
\end{equation}
\end{center}

with $\eta_j = \mu_{j+1}-\mu_{j}, $ for $j > 1$, $\mu_1\leq \mu_2 \leq \ldots \leq \mu_{N}$. Eq (\ref{Mstepstatedep}) can be maximized with regular maximization algorithms. The approximation guarantees the descent property of the EM-algorithm (\citealp{zou_one-step_2008}).\vskip0.4cm
\textit{Non-stationary Markov chain by means of covariates}

When time-varying covariates are included in the hidden process, the parameters to estimate are $\boldsymbol{\Psi}= (\boldsymbol{\delta},\boldsymbol{\beta},\boldsymbol{\mu},\boldsymbol{\sigma})$. The assumption of stationarity no longer holds since the model is time-dependent and the stationary DPMLE described cannot be applied.

Without stationary probabilities, the penalty to prevent overfitting of type I is adjusted, while the rest of the DPMLE and estimation of parameters, except for 
$\boldsymbol{\beta}$, remain unchanged. The stationary distribution is replaced by an approximation $\hat{\boldsymbol{\pi}}$, computed using forward and backward probabilities. At iteration $p+1$, $\boldsymbol{\hat{\beta}}^{(p+1)}$ is updated as follows:
\begin{equation}
\begin{aligned}
\label{newEM}
\boldsymbol{\hat{\beta}}^{(p+1)} &= {\text{argmax}}_{{\boldsymbol{\beta}}}\underset{t=2}{\overset{T}{\sum}}
    \underset{j,i=1}{\overset{N}{\sum}}\hat{v}_{ij}^{(p)}(t)\log{\gamma^{(t)}_{ij}}+ C_N\underset{j=1}{\overset{N}{\sum}}\log\hat{\pi}_j(\hat{\boldsymbol{\delta}}^{(p)},\boldsymbol{\beta},\hat{\boldsymbol{\mu}}^{(p)},\hat{\boldsymbol{\sigma}}^{(p)}), \\
\hat{\pi}_j(\boldsymbol{\Psi}) &= \frac{1}{T}\underset{t=1}{\overset{T}{\sum}}\mathbb{P}[S_t=j|\boldsymbol{Y}=\boldsymbol{y}] = \frac{1}{TL}\underset{t=1}{\overset{T}{\sum}}\alpha_{t}(j)\beta_{t}(j).
\end{aligned}
\end{equation}
We used the following property $\mathbb{P}[S_t=i|\boldsymbol{Y}=\boldsymbol{y}]= \frac{\alpha_{t}(i)\beta_{t}(i)}{L},
$ with likelihood $L$. 

The derivation of $\hat{\boldsymbol{\pi}}$ in Eq (\ref{newEM}) is motivated by the fact that the DPMLE aims at penalizing the proportion of time spent in each state. Therefore, we use estimates of the proportion of time spent in each state as substitutes for stationary probabilities. This method is convenient since a gradient can be computed. The number of operations involved in computing $\hat{\pi}(\boldsymbol{\Psi})$ is of order $TN^2$ (\citealp{zucchini_hidden_2017}). 
The Viterbi algorithm could potentially be used to derive $\hat{\boldsymbol{\pi}}$ (see supporting information) but the gradient cannot be computed, limiting the use of gradient-descent methods to fit the model, and thus is not considered further. 

\cite{manole_estimating_2021} propose the Group-Sort-Fuse (GSF) procedure to adapt
the DPMLE to the multivariate case (i.e., penalty applied to a multidimensional parameter). The GSF generalizes the natural ordering of ${\boldsymbol{\mu}}=({\boldsymbol{\mu}}_1,\ldots, {\boldsymbol{\mu}}_N )$ on the real line to the multivariate space, using $l_2$-norms, and cluster ordering $\tau$ as follows:

\begin{equation}
    \begin{array}{l}
        \boldsymbol{\mu}_{\tau(1)} = \text{argmin}_{\boldsymbol{\mu}_i : i = 1 \ldots N} ||\boldsymbol{\mu}_i||_2 \\[1em]
        \boldsymbol{\mu}_{\tau(k)} = \text{argmin}_{\boldsymbol{\mu}_j \neq \boldsymbol{\mu}_{\tau(i)} \, \forall i} ||\boldsymbol{\mu}_j - \boldsymbol{\mu}_{\tau(k-1)}||_2.
    \end{array}
\end{equation}
$p_{\lambda}$ is then applied to the difference in $l_2$ norms of the clustered parameters.\vskip0.4cm

The choice of the tuning parameters $C_N$ and $\lambda_M$ is crucial to ensure that the method performs well. We set $a =3.7$ (\citealp{hung_hidden_2013}). $\lambda_M$ and $C_N$ are chosen from a discrete set by a BIC-type criterion, denoted Narwhal Information Criterion (NIC), which consists of selecting both hyperparameters that minimize 
\begin{equation}
    \label{BICtypecriterion}
    -2\log \mathcal{L}_M(\hat{\boldsymbol{\Psi}}|\boldsymbol{y})+k\log(n),
\end{equation} 
where $\hat{\boldsymbol{\Psi}}$ is the vector of parameter estimates after the DPMLE procedure (\citealp{wang_tuning_2007}; \citealp{lin_order_2022}). The number of parameters $k$ can be derived as follows:
\begin{center}
\begin{equation}
\label{numparam}
    k=\text{dim}(\boldsymbol{\theta}_1)\widehat{N} + \widehat{N}(\widehat{N}-1),
\end{equation}
\end{center}
where $\text{dim}(\boldsymbol{\theta}_1)$ corresponds to the dimension of $\boldsymbol{\theta}_1$, which essentially represents the number of emission parameters to be estimated for each state. From Eq (\ref{numparam}), $k$ increases with the estimated number states $\widehat{N}$. $\lambda_M$ is expected to increase with the sample size and the choice of $C_N$ is not expected to affect the parameters and order estimates (\citealp{chen_order_2008}; \citealp{hung_hidden_2013}). NIC could potentially share the same limitations as BIC, which is to rely heavily on sample size. However, this should be mitigated by the fact that the DPMLE itself will perform better than BIC and be less reliant on the criterion for selecting the appropriate hyperparameters and thus number of states. 

\subsection*{Simulation study}
\label{Simulation}
We implemented a simulation framework based on \cite{pohle_selecting_2017}. We use similar scenarios, each exploring a type of complexity commonly found in animal movement data: (1) benchmark (no misspecification), (2) outliers, (3) individual heterogeneity in the hidden process, (4) individual heterogeneity in the observed process, (5) violation of the conditional independence assumption and (6) temporal variation in the hidden process. 

\begin{figure}[htbp]
     \hspace{-0.5cm}
    \begin{subfigure}[b]{0.33\textwidth}  
    \includegraphics[width=1\textwidth]{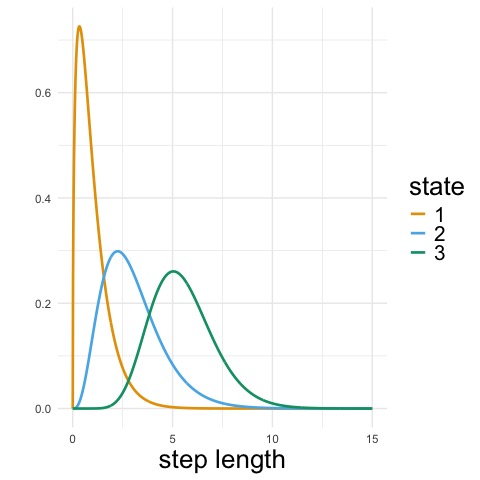}
        \caption{\centering Gamma densities for the benchmark model.}
        \label{scenario1}
    \end{subfigure}
    \begin{subfigure}[b]{0.33\textwidth}
        \hspace{0.2cm}
\includegraphics[width=\textwidth]{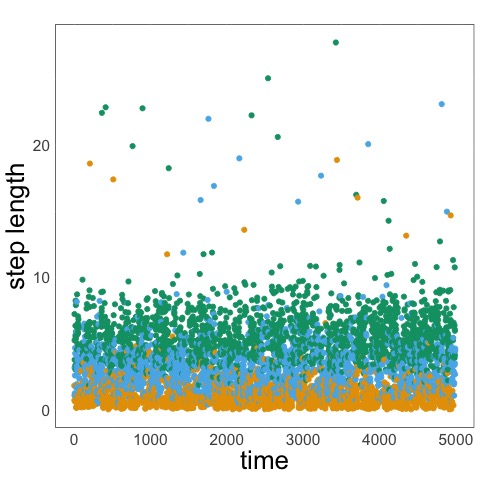}
        \caption{\centering Example of time series generated in scenario 2.}
        \label{scenario2}
    \end{subfigure}
        \hspace{1.2cm}
    \begin{subfigure}[b]{0.33\textwidth}    
\includegraphics[width=1.05\textwidth]{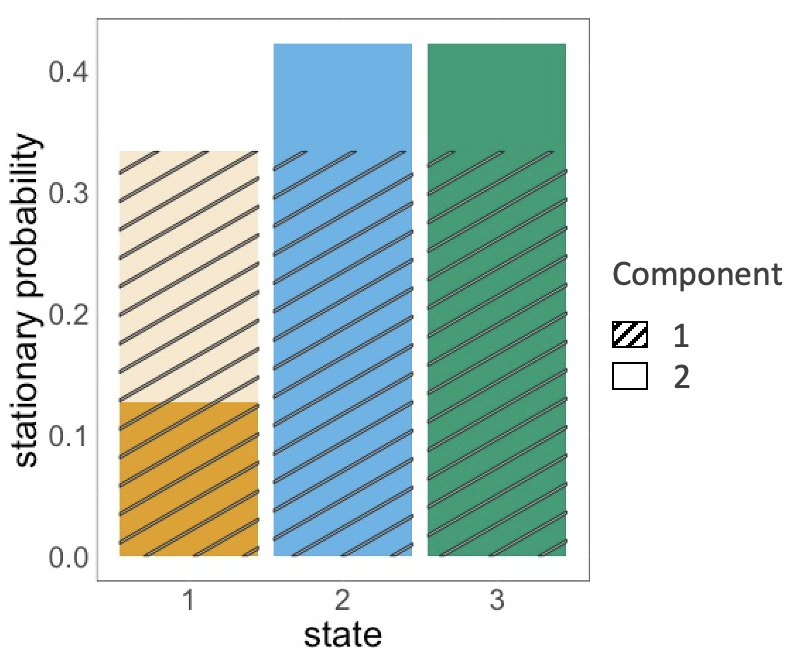}
        \caption{\centering Stationary probabilities in both components in scenario 3.}
        \label{scenario3}
    \end{subfigure}
    \vskip0.75cm
         \hspace{-0.5cm}
      \begin{subfigure}[b]{0.33\textwidth}
\includegraphics[width=1\textwidth]{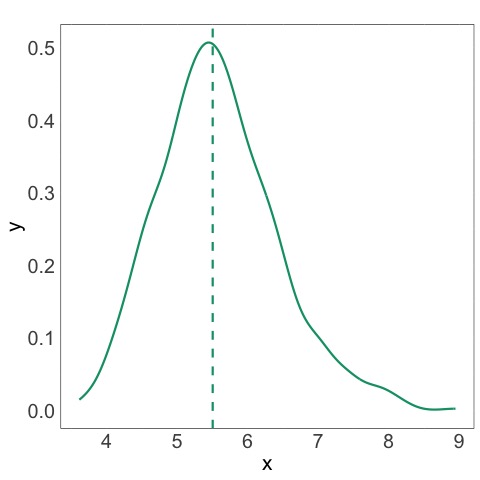}
        \caption{\centering Log-normal density for the mean values of state $3$ in scenario 4.}
        \label{scenario4}
    \end{subfigure}
    \begin{subfigure}[b]{0.33\textwidth}
            \hspace{0.2cm}
\includegraphics[width=1\textwidth]{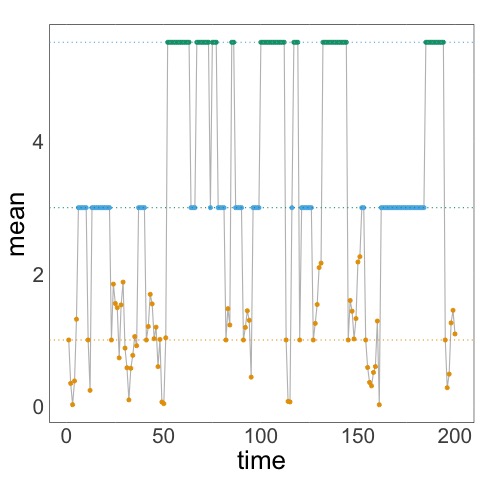}
     \caption{ \centering Example sequence of gamma mean values generated in Scenario 5.}
        \label{scenario5}
    \end{subfigure}
        \hspace{1.2cm}
    \begin{subfigure}[b]{0.33\textwidth}    \includegraphics[width=\textwidth]{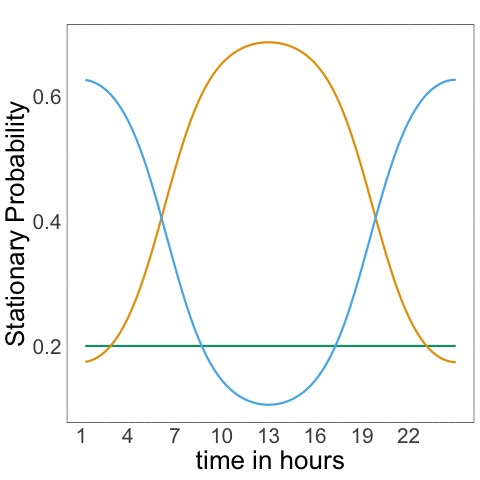}
        \caption{\centering Stationary probabilities over time in Scenario 6.}
        \label{scenario6}
    \end{subfigure}
    \caption[]{\label{scenarios}Simulation scenarios: state 1 (orange), state 2 (blue), and state 3 (green).}
    \label{rfidtag_testing}
\end{figure}

We simulate three-state HMMs across all scenarios. 
The baseline model used for all scenarios is a stationary gamma–HMM (Eq (\ref{nullmulti}) with gamma state-dependent distributions) to which additional structure (under different scenarios) is incorporated  (Fig$.$ \ref{scenarios}). Scenarios 1, 2 and 6 generate the time series of only one individual, while scenarios 3, 4 and 5 use the same parameters to generate a time series for ten independent individuals. We explore two sample sizes:  $T=5,000$ and $T=12,000$. 

Scenario 1 is defined to show the performance of the methods without misspecification (Fig$.$ \ref{scenario1}). The data are generated following a standard three-state stationary gamma–HMM with means $(\mu_1,\mu_2,\mu_3) = (1,3,5.5)$ and shapes $(s_1,s_2,s_3)=(1.5,4,12)$. The tpm is 

$$
\Gamma = 
\begin{pmatrix}
  0.8 & 0.1 & 0.1 \\
0.1 & 0.8 & 0.1 \\ 
0.1 & 0.1 & 0.8
\end{pmatrix}.
$$

These mean and shape values were selected because the ICL criterion outperforms BIC and AIC for most scenarios, but performs particularly poorly in selecting the right number of states under this setting (\citealp{pohle_selecting_2017}). Scenario 2 incorporates outliers in the observations by adding uniformly distributed random errors from the interval $[10, 20]$ to 0.5\% of the data simulated with the framework of scenario 1 (\citealp{pohle_selecting_2017}). To include inter-individual differences (e.g., stressed/baseline individuals, male/female), scenario 3 adds discrete random effects in the hidden process. We simulate individuals with equal time series length $T_m = T,$ for all $ m \leq 10$. Discrete random effects in the hidden process assume that the time series are clustered in $K>1$ components, where each component has its tpm and the members of the same component share a common set of parameters (\citealp{deruiter_multivariate_2017}; \citealp{mcclintock_worth_2021}). We consider $K=2$ with likelihood:
\begin{center}
\begin{equation}
\mathcal{L}_{mix} = \underset{m=1}{\overset{M}{\prod}}\underset{k=1}{\overset{K}{\sum}}\delta^{(k)}\boldsymbol{\Gamma}^{(k)} \textbf{P}(\boldsymbol{y}^{m}_{1})\boldsymbol{\Gamma}^{(k)} \textbf{P}(\boldsymbol{y}^{m}_{2})\ldots\boldsymbol{\Gamma}^{(k)} \textbf{P}(\boldsymbol{y}^{m}_{T_m})\textbf{1}\nu^{(k)}, 
\end{equation}
\end{center}
 with $
\boldsymbol{\Gamma}^{(1)}
$ as in scenario 1
and 
$
\boldsymbol{\Gamma}^{(2)}=
\begin{pmatrix}
  0.1 & 0.1 & 0.8 \\
  0.1 & 0.8 & 0.1 \\ 
  0.1 & 0.1 & 0.8
\end{pmatrix},
$
where $\boldsymbol{\nu}=(\nu^{(1)},\nu^{(2)})=(0.5,0.5)$ is the vector of mixture probabilities. To include individual heterogeneity in the observation process (e.g., juveniles may be slower than adults; \citealp{demars_inferring_2013}), we follow the procedure defined by \cite{pohle_selecting_2017} and generate the individual means of the gamma emission distribution within the third state (Fig$.$ \ref{scenario4}) with a log-normal distribution with mean and variance parameters $(\log(5.5),0.15)$. Scenario 5 includes additional correlation in the observational process. We generate individuals with time-varying mean parameters for the first state (Fig$.$ \ref{scenario5}), simulated using an auto-regressive process of order $1$ with persistence $0.85$. Scenario 6 incorporates temporal variation in the transition probabilities (e.g., foraging is more likely at night) with the cosinor function and a resolution of $15$ minutes (Fig$.$ \ref{scenario6}).\vskip0.4cm

All scenarios, except scenario $5$, are simulated with the \texttt{simData} function from the \texttt{momentuHMM} package (\citealp{mcclintock_momentuhmm_2018}) in \texttt{R} (\citealp{R_2021}). We fit both stationary and non-stationary approaches, using the covariate "time of day" for the non-stationary methods and omitting it for their stationary versions, resulting in a total of eight models to fit for each scenario. In contrast to scenario 6, the covariate "time of the day" used in the non-stationary models fitted is linear ($1$ to $96$).

To select the best model according to AIC and BIC, we fit standard stationary gamma–HMMs with $2,3$ and $4$ states to each simulated dataset and select the model minimizing each criterion. For each model fitted, we explore $150$ random initial values (\citealp{pohle_selecting_2017}), and select the parameter estimates corresponding to the minimum negative joint log-likelihood. For the DPMLE methods, we use MLE estimates from random positions that maximized the observed log-likelihood as initial values and explore nine random initial values. Fewer random initial values are explored compared to AIC and BIC since the method takes longer to converge and has MLE estimates for starting parameters, allowing for efficient convergence despite the reduced exploration. We perform random search for $\lambda_{M}$ and $C_N$, along with NIC to select the hyperparameters. We explore 50 uniformly distributed values of $\lambda_{M}$ and $C_N$, and set the interval for $\log M\lambda_{M}$ and $C_N$ to $[1,5]$. We do not explore $C_N$ in the log scale since the penalized likelihood method is not sensitive to the choice of $C_N$ (\citealp{hung_hidden_2013}; \citealp{lin_order_2022}). 

For each scenario and sample size, $100$ datasets are generated, and the success rates (percentage (\%) of correctly estimated number of states) of each method are compared. In total six scenarios with two different sample sizes, representing $1200$ different datasets are generated. This corresponds to $60000$ simulation runs per DPMLE method. The upper bound for the number of states across every method is set to $N=4$ for computational feasibility of the simulation framework. 

\subsection*{Narwhal case study}
\label{narval}
\begin{figure}[H]
    \hspace{1cm}
    \includegraphics[width=0.75\textwidth]{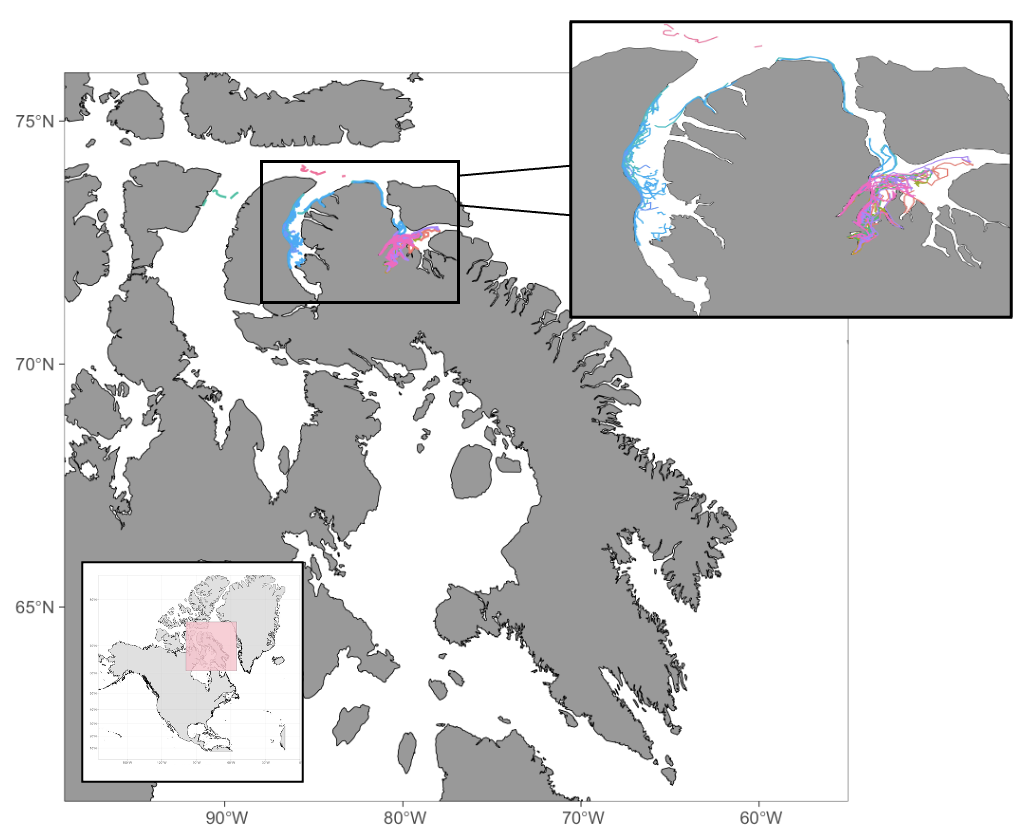}
        \caption{Map showing the location
    data for eight narwhal tracked from August 2017 to October 2017, after data cleaning. Each colour corresponds to a different individual track.}
        \label{narvalmap}
\end{figure}

We demonstrate the use of our method on a narwhal case study. The main goal is to estimate the number of behavioural states given narwhal movement patterns and understand their relationship with environmental covariates. The study focuses on the region of Qikiqtaaluk (Baffin) in Nunavut, Canada. During the summer 2017, $18$ narwhal were equipped with electronic tags in Tremblay Sound (72°30N,80°45W; protocols for narwhal's capture and tagging are described in \citealp{shuert_decadal_2022}). We will demonstrate the performance of the proposed method for order selection using location data only for eight narwhal with FastLoc GPS data for a period of two months with a temporal resolution of an hour (Fig$.$ \ref{narvalmap}). More details regarding data processing can be found in the supporting information. Latitude and longitude were converted into step length and turning angle. Step lengths were modelled with gamma distributions and turning angle with von Mises distributions. The GSF procedure was used to adapt the DPMLE methods to the multivariate case (i.e., overlapping states were characterized by vectors of mean step length and turning angle concentrations "close" to each other).

We did not expect the number of states $N^{\star}$ to exceed $4$ (\citealp{pohle_selecting_2017}; \citealp{ngo_understanding_2019}), thus we set the upper bound to be $N = 8$ to allow $N > N^{\star}$. We
explored a hundred pairs of randomly sampled hyperparameters $\lambda_M$ and $C_N$ and selected the best pair with NIC (Eq \ref{BICtypecriterion}). We set $a= 3.7$. We investigated the covariate "distance to shore" (\citealp{breed_sustained_2017}; \citealp{shuert_divergent_2023}). BIC was used to choose between DPMLE models with and without the covariate. We fitted standard HMMs with momentuHMM using $30$ random starting values and initialized both DPMLE methods with MLE estimates. Due to computational constraints, only $10$ random starting values were explored for the DPMLE methods.

\section{Results}
\label{Res}
\subsection*{Simulated data}
\label{ResSim}
\begin{figure}[H]
    \hspace{-2cm}   
    \includegraphics[width=1.3\linewidth]{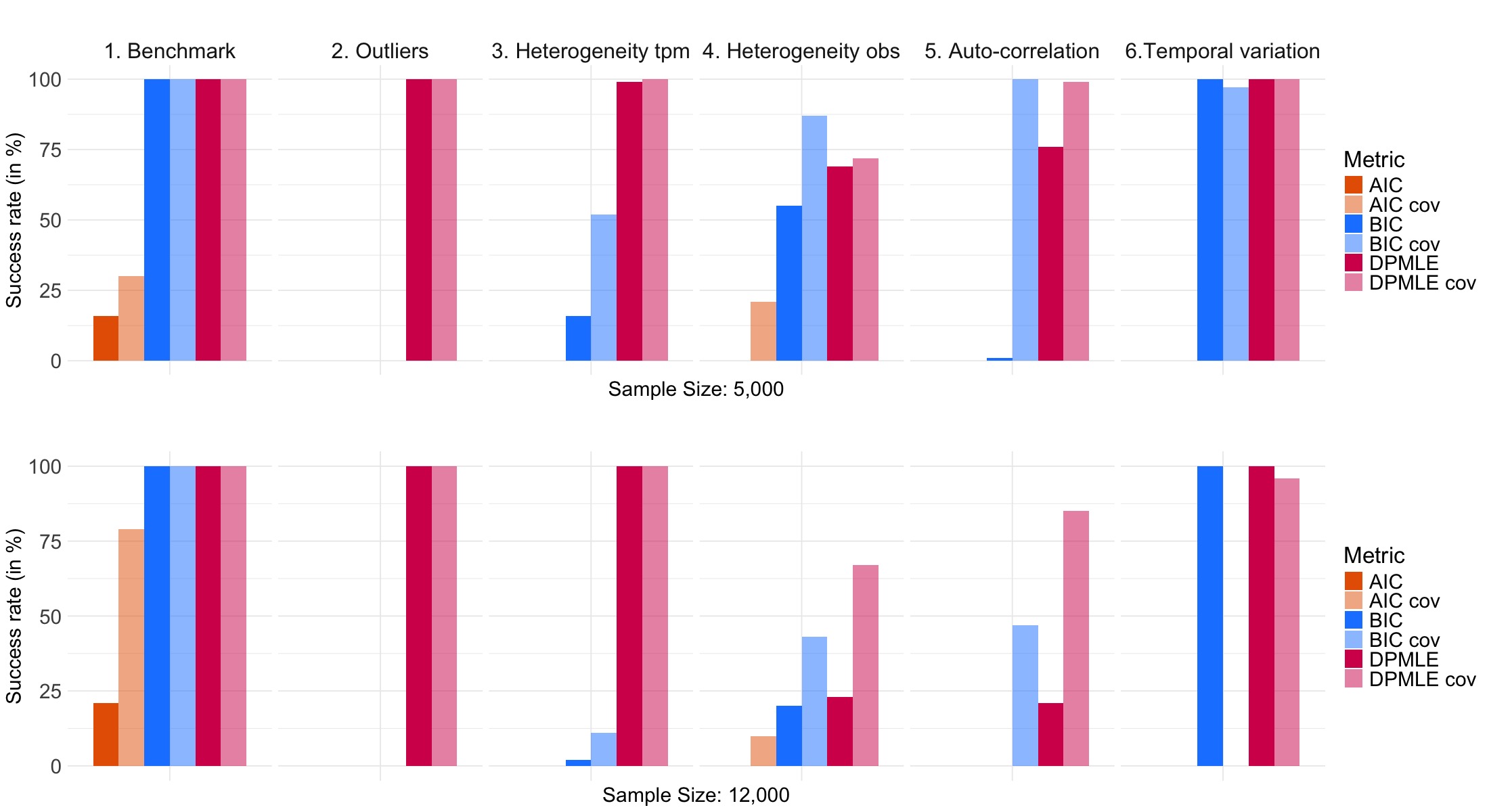}
    \caption{Success rate in estimating the correct number of states (i.e., three) for each scenario, with different sample sizes. "cov" refers to models fitted with the linear covariate "time of day".}
    \label{fig:res}
\end{figure}

Both DPMLE methods outperformed BIC and AIC, and had a success rate higher than $99\%$ for more than half of the scenarios (Fig$.$ \ref{fig:res}). The non-stationary DPMLE was consistently among the best-performing methods and had more than $85\%$ success rate for ten out of the twelve simulation settings explored. The mean estimates of both DPMLE methods closely aligned with the simulated values (Figs \ref{boxplots}–\ref{boxplotsbenchmark}).
 BIC performed well for some scenarios, and performed better than the DPMLE in two cases (Scenario 4 and Scenario 5 with covariates with a sample size of $5,000$; Fig$.$ \ref{fig:res}). However, its performance was significantly reduced when applied to a large sample size ($12,000$). The performance of the non-stationary DPMLE was generally much higher and was less affected by an increase in sample size than AIC and BIC. 

Under scenario $2$ (presence of outliers), both DPMLEs identified the correct number of states with a $100$\% success rate, while BIC and AIC consistently overestimated the number of states (see tables in supporting information). The DPMLEs also exhibited strong performance under scenarios $3$ (heterogeneity in the tpm) and $6$ (temporal variation in tpm) with more than $90\%$ success rates for both sample sizes, which is likely because the misspecification is in the tpm rather than the state-dependent distributions. Both double penalized likelihood methods estimate $N(N-1)$ transition probabilities, and can therefore incorporate the misspecification in the additional states. The tpm for merged states can be estimated by averaging the state-transition probabilities (see supporting information). Both DPMLE methods tended to slightly underestimate the means of the last two states under scenario $2$ and $3$, with accuracy improving as the sample size increased (see Figs \ref{boxplots}–\ref{boxplotsbenchmark}). Non-stationary DPMLE mean estimates under scenario $6$ outperformed the stationary estimates as the sample size increased.

In scenario $4$ and $5$ with sample size of $5,000$, BIC with covariates performed better than both DPMLEs. However, its performance was halved when the sample size increased to $12,000$. This suggests that the inclusion of covariates improves the performance of BIC through the increase of its penalty value, leading to the selection of fewer states. As the sample size increases, the likelihood becomes the dominant factor, which explains the decrease in performance (i.e., increase in overfitting) observed with higher sample sizes. A similar behaviour arose with both DPMLEs for the same scenarios. However, the decrease in performance was only drastic for the stationary method, and the non-stationary DPMLE still performed reasonably well under scenario $5$ with an $85\%$ success rate for a sample size of $12,000$. Under scenario $5$, AIC selected an additional state between state $2$ and $3$ to incorporate the misspecification. BIC behaved similarly $53\%$ of the time for a sample size of $12,000$. The non-stationary DPMLE method was less subject to this behaviour. Instead, it slightly underestimated the mean estimates (Fig$.$ \ref{boxplots} in supporting information) and incorporated some of the temporal variation in the temporal covariate.\vskip0.4cm

\subsection*{Narwhal movement data}

Both DPMLE procedures selected two states, whereas model selection criteria selected more than four states. The non-stationary DPMLE was selected as the best performing DPMLE model. BIC identified the five-state HMM without covariate as the best-performing model. As expected from the simulations, AIC selected the most complex model available: eight states with covariate "distance to shore". 
\begin{figure}[htbp]
 \begin{subfigure}[b]{0.33\textwidth}  
     \centering
    \includegraphics[width=0.85\textwidth]{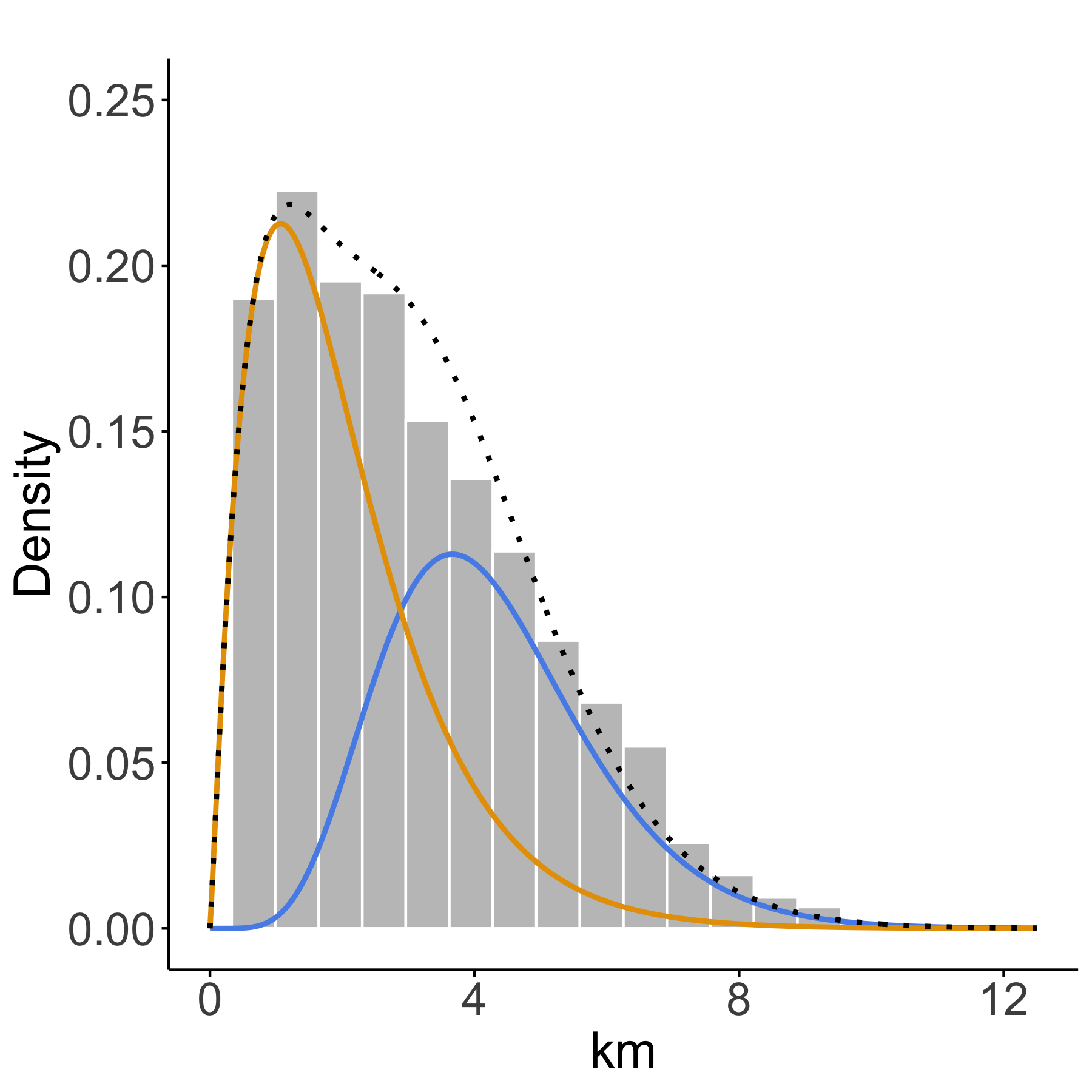}
        \label{}
        \caption{\centering Step length histograms and associated estimated densities.}
    \end{subfigure}
      \begin{subfigure}[b]{0.33\textwidth}
          \centering
\includegraphics[width=0.85\textwidth]{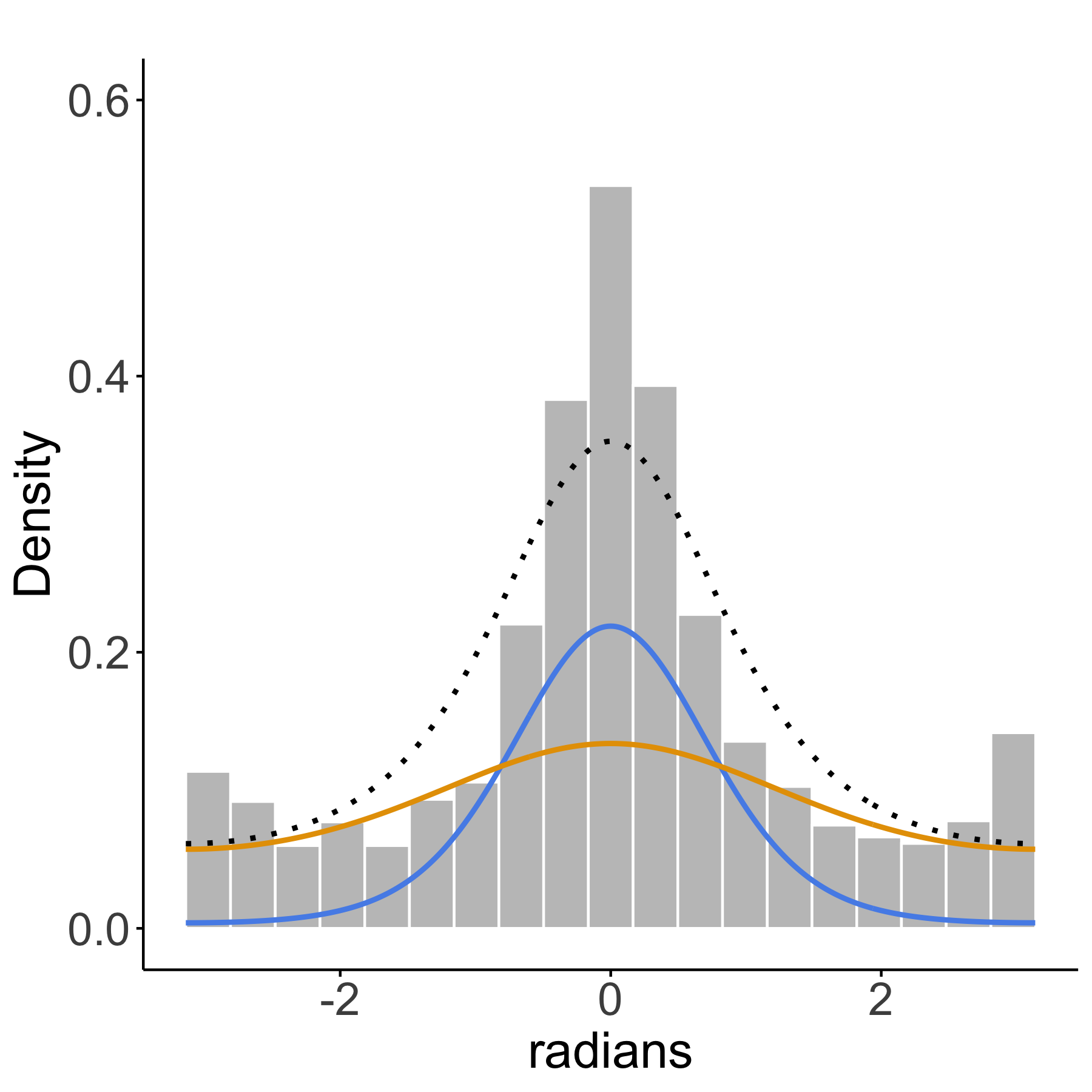}
\caption{\centering Turning angle histograms and associated estimated densities.}
        \label{}
    \end{subfigure}
       \begin{subfigure}[b]{0.33\textwidth}
                  \centering
\includegraphics[width=0.85\textwidth]{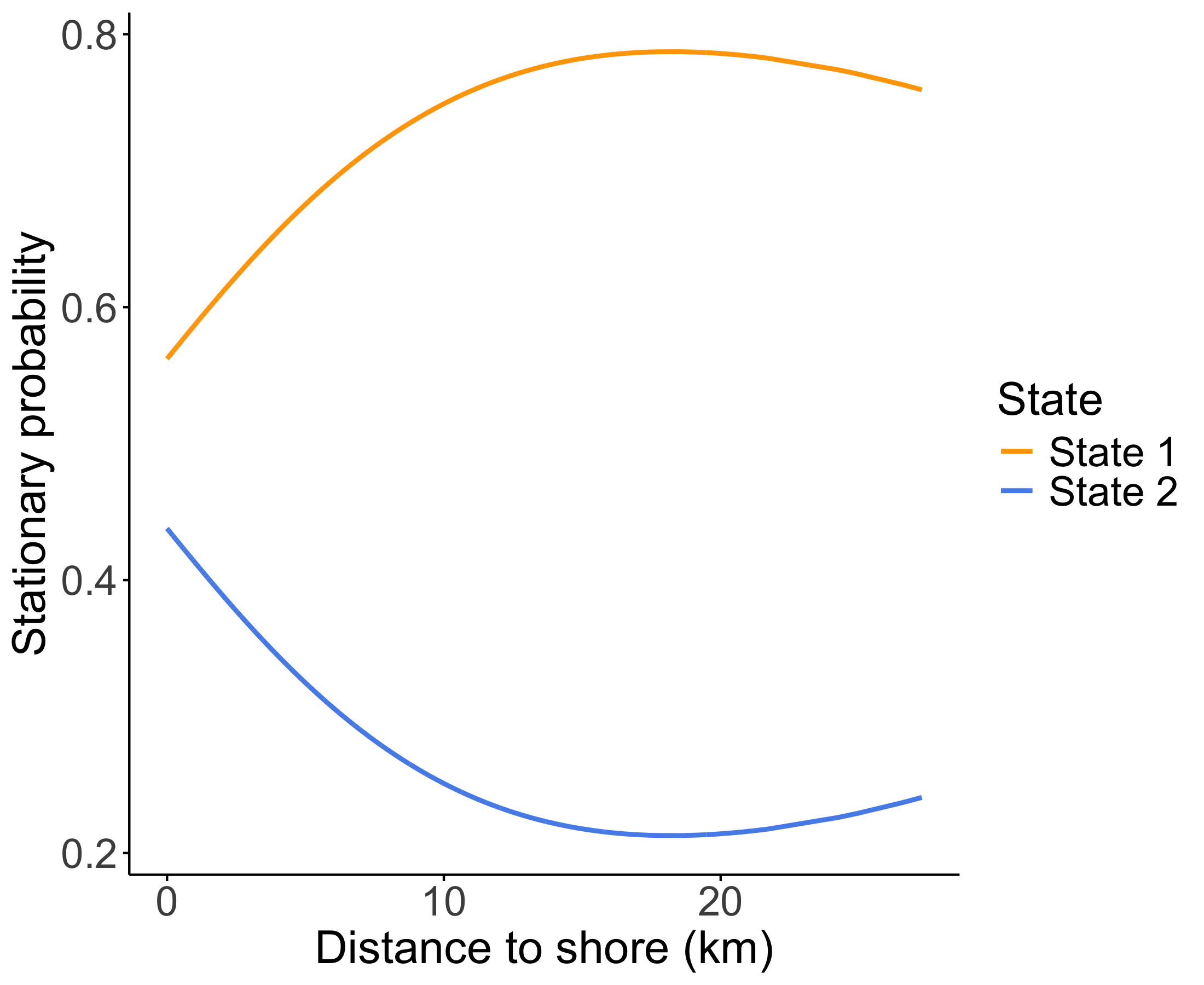}
        \caption{\centering Estimated "stationary" probabilities.}
    \end{subfigure}
    \caption{\label{resultsDPMLE} 
Estimates from the HMM of narwhal movement data obtained with the non-stationary DPMLE, with the two estimated states (in blue and orange).}
    \end{figure}
    
\begin{figure}[H] 
  \begin{flushleft}
      \begin{subfigure}[b]{0.45\textwidth}
       \centering 
        \includegraphics[width=0.6\textwidth]{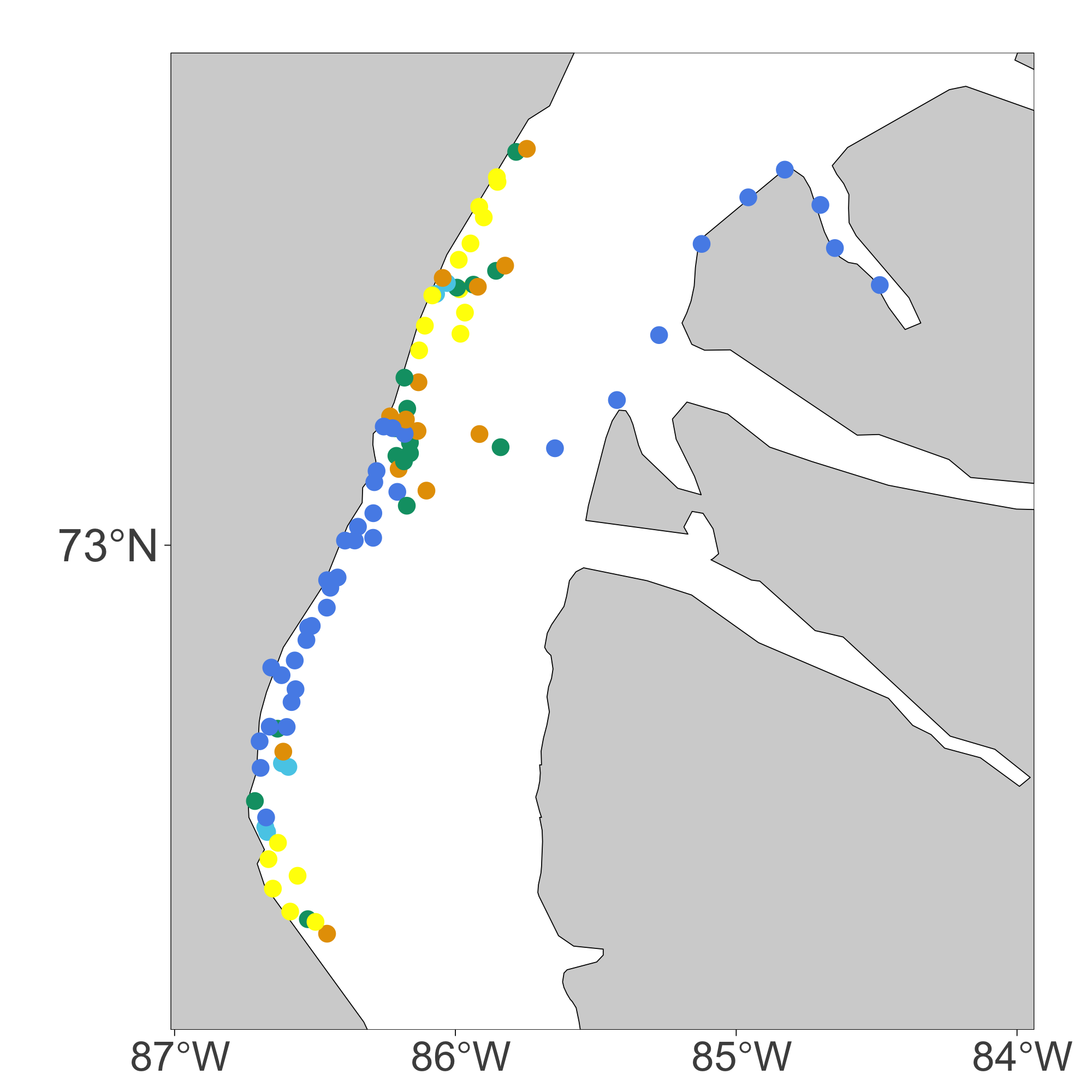}
        \label{tracksBIC}
             \caption{}

    \end{subfigure}
    \begin{subfigure}[b]{0.45\textwidth}
\includegraphics[width=0.6\textwidth]{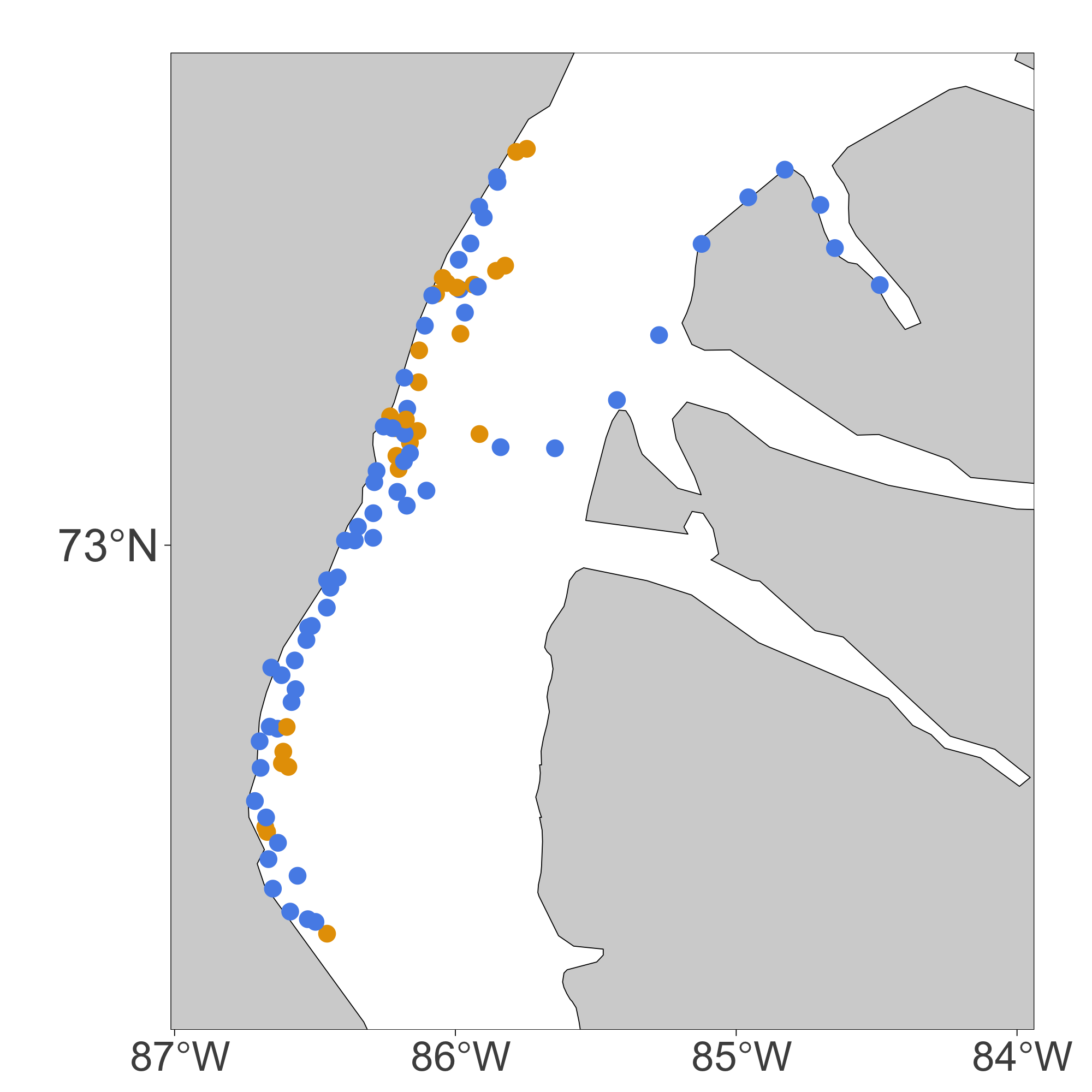}
        \label{tracksDPMLE}
             \caption{}

    \end{subfigure}
\vskip0.7cm
      \begin{subfigure}[b]{0.45\textwidth}
        \hspace{1.75cm} 
\includegraphics[width=0.6\textwidth]{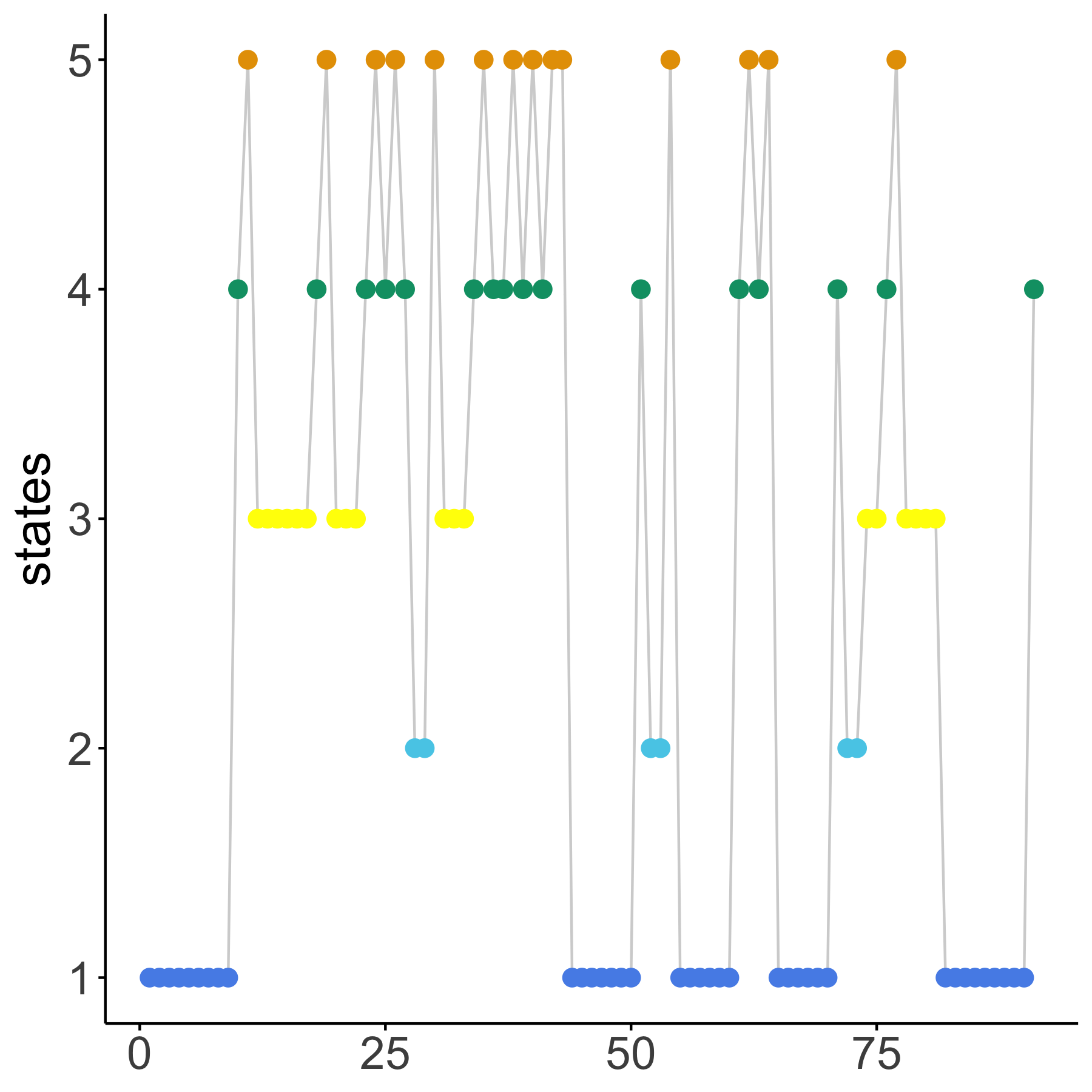}
     \caption{}

        \label{statesBIC}
    \end{subfigure}
    \begin{subfigure}[b]{0.45\textwidth}
       \hspace{0.5cm} 
\includegraphics[width=0.6\textwidth]{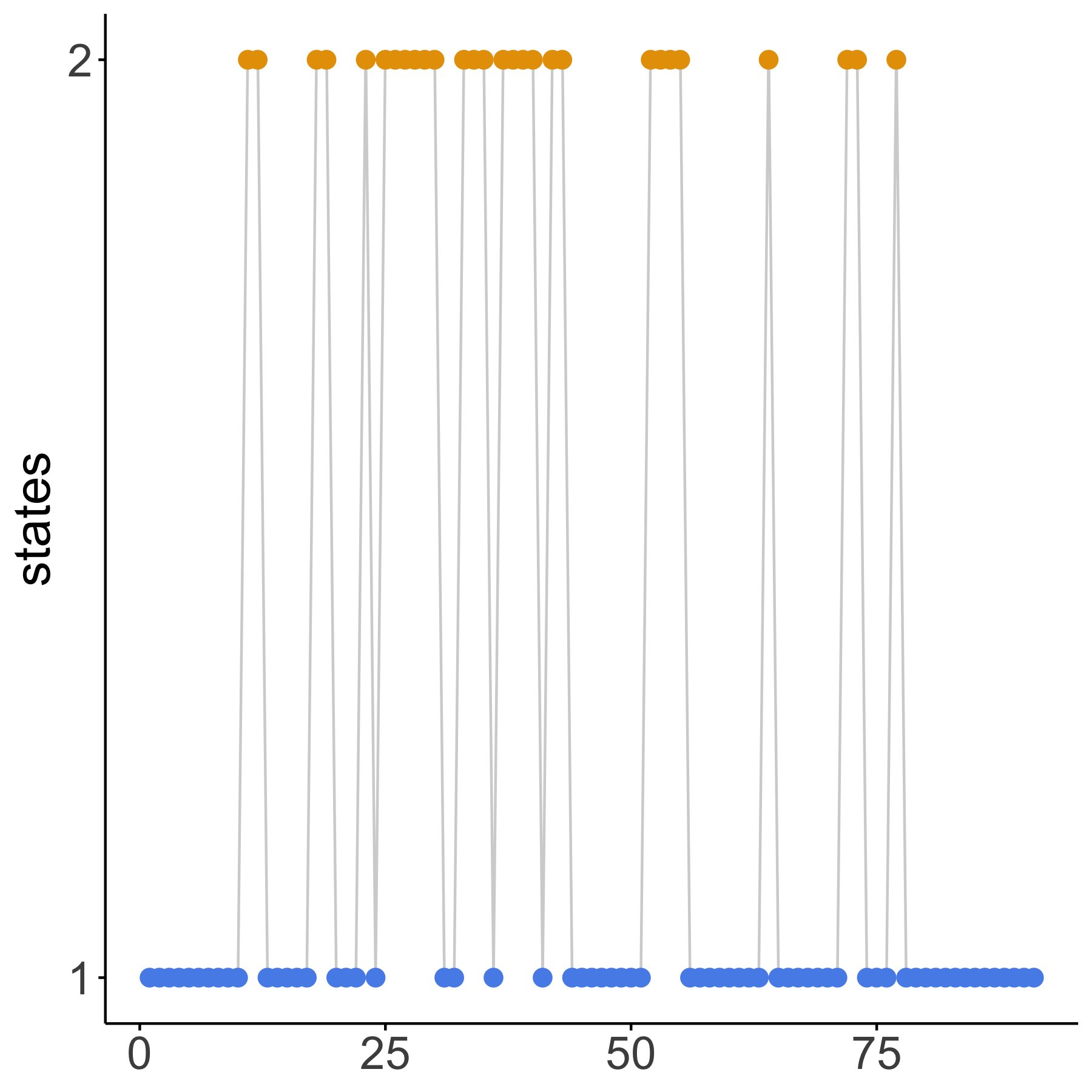}
     \caption{}
        \label{statesDPMLE}
    \end{subfigure}
\caption{\label{MPLEfit}Narwhal locations coloured by state from HMM model selected by BIC (a) and non-stationary DPMLE (b) with associated state time-series (c and d, respectively).}
 \end{flushleft}
\end{figure}

The agreement between both DPMLE methods provided strong evidence that narwhal movement in the summer is best explained by two behavioural states. Movement of type $1$ (Fig$.$ \ref{resultsDPMLE}) estimated by the 2-state non-stationary DPMLE was slow and with little directional persistence, thus interpreted as area-restricted searching behaviour, which is typically characterized by the ability to adjust movement adaptively (\citealp{dorfman2022guide}). This state can cover different behaviours such as foraging, resting and socializing activities. Movement of type $2$ was faster and with strong directional persistence which likely reflected a transiting behaviour. This interpretation was corroborated by the decoded states obtained from the forward-backward algorithm (Fig$.$ \ref{MPLEfit}). Extreme turning angle values falling outside the estimated distribution were interpreted as location errors (Fig$.$ \ref{resultsDPMLE}; \citealp{hurford_gps_2009}).
The states of the two-state HMM were stable, as shown by their high persistence and the decoded state sequence (Fig$.$ \ref{statesDPMLE}). On average, both states exhibited high persistence ($0.88$ and $0.81$). From Fig$.$ \ref{resultsDPMLE},
time spent in slow movement increased further from the shoreline, while time spent in the directed state increased as narwhal approach the shoreline. Narwhal typically feed on deep-water prey, which could explain the association between deeper waters (i.e., areas far from the shoreline) and increased time spent in area-restricted searching behaviour, as these areas are likely rich in food resources (\citealp{watt_fatty_2015}). In contrast, the increased time spent in the directed state near the shoreline may represent an adaptive anti-predatory response to the rising threat of killer whale predation in the Arctic (\citealp{breed_sustained_2017}).

In the five-state model, states did not have a clear interpretation and exhibited overlapping step length distributions (Fig$.$ \ref{BICfit} in supporting information). Only one state could be identified as a transiting behaviour (blue). States $4$ and $5$ had very low persistence ($\gamma_{11}=0.17$, $\gamma_{55}=0.25$) and seemed unstable (Fig$.$ \ref{statesBIC}). Thus, they were difficult to interpret as distinct and well-defined behaviours. Strong oscillations between states could indicate the presence of additional and unnecessary states (\citealp{celeux_selecting_2008}). In contrast, two-state HMMs differentiating between a resident (i.e., area-restricted search) and a transiting (fast and directed) movement are very common in animal movement analyses (\citealp{whoriskey_hidden_2017}). Movement associated with the turquoise, orange and green states from the $5$-state HMM (Fig$.$ \ref{BICfit} in supporting information) could correspond to area-restricted searching behaviour estimated by the $2$-state HMM (characterized by low directional persistence and slower movement) and the other two states to transiting.

\section{Discussion}
\label{discussion}
We proposed a double penalized maximum likelihood estimate to perform simultaneous parameter estimation and order selection in non-stationary HMMs that overcame most of the problems associated with AIC and BIC. In the simulation study, our method significantly outperformed both criteria. Our narwhal movement case study demonstrated that, unlike AIC and BIC, our proposed method identified two behavioural states that closely aligned with the expected movement behaviour for this species (\citealp{watt_fatty_2015}; \citealp{breed_sustained_2017}; \citealp{shuert_divergent_2023}), further demonstrating its usefulness in ecology. The non-stationary DPMLE has demonstrated greater efficiency in handling the narwhal movement data compared to the stationary DPMLE and allowed to model a relationship with an important environmental covariate.

DPMLE is particularly effective at handling misspecification in the hidden process (scenarios 3 and 6) because it estimates a large tpm that can accommodate model misspecification. With a higher upper bound, the method is expected to handle greater levels of misspecification. However, merged states may exhibit distinct transition probabilities, which must not be overlooked. DPMLE cannot handle large levels of heterogeneity in emission distributions ($23$\%-$67$\% success rate in scenario $4$ with $T = 12,000$) likely because such heterogeneity creates distinct means that are ascribed to different states. Individual variation in movement between animals is inherent in nature but we would expect the performance of both DPMLE methods to improve as differences between individuals decrease. One possible solution to circumvent this inherent problem is to fit one model per individual. Several misspecifications were not explored here, such as those arising from inadequate emission distributions (\citealp{de_chaumaray_estimation_2022}). 

Information criteria are unreliable for order selection in HMMs applied to complex real-world data. AIC consistently overestimated the number of states, while BIC showed better performance under certain conditions. However, for most misspecified scenarios, BIC tended to favour models with too many states. In the case of the narwhal dataset, both criteria selected many states that were difficult to interpret, further illustrating their limitations when applied to complex real-world data and in agreement with the findings of previous studies (\citealp{pohle_selecting_2017}; \citealp{li_incorporating_2017}).

While the simulation results show the good performance of the proposed method with various sample sizes, the non-stationary DPMLE is computationally expensive and future work formally assessing its asymptotic properties is required. We  also note that NIC is a BIC-type criterion that relies on the effective number of parameters and has only been rigorously derived in Gaussian models (\citealp{fan_variable_2001}; \citealp{wang_tuning_2007}). 
Cross-validation could be an alternative approach for hyperparameter selection method, but the computational costs of the DPMLE currently make this option impractical. Additionally, the potential overfitting issues highlighted by \cite{wang_tuning_2007} warrant further investigation within the HMM framework.

We have introduced a DPMLE that, for the first time, provides a flexible framework for handling non-stationary HMMs, thereby expanding its use to a wider range of real-life scenarios compared to other methods. While it is important to minimize misspecification by modelling as much relevant information as possible, movement ecologists rarely have the data or knowledge needed to create a fully specified model. Due to its improved performance in the simulation study and when applied to real-world data, the non-stationary DPMLE provides a good surrogate to standard information criteria to perform order selection and parameter inference simultaneously while handling the remaining misspecifications in the data. We recommend using a non-stationary DPMLE with the largest feasible upper bound (considering computational constraints) and using a similar set of bounds to the one used in the simulations for the hyperparameters.

\newpage
\bibliographystyle{chicago}
\bibliography{bibli}

\begin{thebibliography}{}

\bibitem[\protect\citeauthoryear{Akaike}{Akaike}{1974}]{akaike_new_1974}
Akaike, H. (1974).
\newblock A new look at the statistical model identification.
\newblock {\em IEEE Transactions on Automatic Control\/}~{\em 19\/}(6), 716--723.

\bibitem[\protect\citeauthoryear{Auger‐Méthé, Newman, Cole, Empacher, Gryba, King, Leos‐Barajas, Mills~Flemming, Nielsen, Petris, and Thomas}{Auger‐Méthé et~al.}{2021}]{augermethe_guide_2021}
Auger‐Méthé, M., K.~Newman, D.~Cole, F.~Empacher, R.~Gryba, A.~A. King, V.~Leos‐Barajas, J.~Mills~Flemming, A.~Nielsen, G.~Petris, and L.~Thomas (2021).
\newblock A guide to state–space modeling of ecological time series.
\newblock {\em Ecological Monographs\/}~{\em 91\/}(4).

\bibitem[\protect\citeauthoryear{Baum, Petrie, Soules, and Weiss}{Baum et~al.}{1970}]{baum_maximization_1970}
Baum, L.~E., T.~Petrie, G.~Soules, and N.~Weiss (1970).
\newblock A maximization technique occurring in the statistical analysis of probabilistic functions of {Markov} chains.
\newblock {\em The Annals of Mathematical Statistics\/}~{\em 41\/}(1), 164--171.

\bibitem[\protect\citeauthoryear{Biernacki, Celeux, and Govaert}{Biernacki et~al.}{2000}]{biernacki_assessing_2000}
Biernacki, C., G.~Celeux, and G.~Govaert (2000).
\newblock Assessing a mixture model for clustering with the integrated completed likelihood.
\newblock {\em IEEE Transactions on Pattern Analysis and Machine Intelligence\/}~{\em 22\/}(7), 719--725.

\bibitem[\protect\citeauthoryear{Bolker, Brooks, Clark, Geange, Poulsen, Stevens, and White}{Bolker et~al.}{2009}]{bolker_generalized_2009}
Bolker, B.~M., M.~E. Brooks, C.~J. Clark, S.~W. Geange, J.~R. Poulsen, M.~H.~H. Stevens, and J.-S.~S. White (2009).
\newblock Generalized linear mixed models: a practical guide for ecology and evolution.
\newblock {\em Trends in Ecology \& Evolution\/}~{\em 24\/}(3), 127--135.

\bibitem[\protect\citeauthoryear{Breed, Matthews, Marcoux, Higdon, LeBlanc, Petersen, Orr, Reinhart, and Ferguson}{Breed et~al.}{2017}]{breed_sustained_2017}
Breed, G.~A., C.~J.~D. Matthews, M.~Marcoux, J.~W. Higdon, B.~LeBlanc, S.~D. Petersen, J.~Orr, N.~R. Reinhart, and S.~H. Ferguson (2017).
\newblock Sustained disruption of narwhal habitat use and behavior in the presence of {Arctic} killer whales.
\newblock {\em Proceedings of the National Academy of Sciences\/}~{\em 114\/}(10), 2628--2633.

\bibitem[\protect\citeauthoryear{Celeux and Durand}{Celeux and Durand}{2008}]{celeux_selecting_2008}
Celeux, G. and J.-B. Durand (2008).
\newblock Selecting hidden {Markov} model state number with cross-validated likelihood.
\newblock {\em Computational Statistics\/}~{\em 23\/}(4), 541--564.

\bibitem[\protect\citeauthoryear{Chen and Khalili}{Chen and Khalili}{2008}]{chen_order_2008}
Chen, J. and A.~Khalili (2008).
\newblock Order selection in finite mixture models with a nonsmooth penalty.
\newblock {\em Journal of the American Statistical Association\/}~{\em 103\/}(484), 1674--1683.

\bibitem[\protect\citeauthoryear{Chen, Tan, and Zhang}{Chen et~al.}{2008}]{chen_consistency_2005}
Chen, J., X.~Tan, and R.~Zhang (2008).
\newblock Consistency of penalized mle for normal mixtures in mean and variance.
\newblock {\em Statistica Sinica\/}~{\em 18}, 443--465.

\bibitem[\protect\citeauthoryear{Dannemann and Holzmann}{Dannemann and Holzmann}{2008}]{dannemann_testing_2008}
Dannemann, J. and H.~Holzmann (2008).
\newblock Testing for two states in a hidden {Markov} model.
\newblock {\em Canadian Journal of Statistics\/}~{\em 36\/}(4), 505--520.

\bibitem[\protect\citeauthoryear{de~Chaumaray, Kolei, Etienne, and Marbac}{de~Chaumaray et~al.}{2022}]{de_chaumaray_estimation_2022}
de~Chaumaray, M. D.~R., S.~E. Kolei, M.-P. Etienne, and M.~Marbac (2022).
\newblock Estimation of the order of non-parametric hidden markov models using the singular values of an integral operator.
\newblock {\em arXiv preprint arXiv:2210.03559\/}.

\bibitem[\protect\citeauthoryear{DeMars, Auger‐Méthé, Schlägel, and Boutin}{DeMars et~al.}{2013}]{demars_inferring_2013}
DeMars, C.~A., M.~Auger‐Méthé, U.~E. Schlägel, and S.~Boutin (2013).
\newblock Inferring parturition and neonate survival from movement patterns of female ungulates: a case study using woodland caribou.
\newblock {\em Ecology and Evolution\/}~{\em 3\/}(12), 4149--4160.

\bibitem[\protect\citeauthoryear{DeRuiter, Langrock, Skirbutas, Goldbogen, Calambokidis, Friedlaender, and Southall}{DeRuiter et~al.}{2017}]{deruiter_multivariate_2017}
DeRuiter, S.~L., R.~Langrock, T.~Skirbutas, J.~A. Goldbogen, J.~Calambokidis, A.~S. Friedlaender, and B.~L. Southall (2017).
\newblock A multivariate mixed hidden {Markov} model for blue whale behaviour and responses to sound exposure.
\newblock {\em The Annals of Applied Statistics\/}~{\em 11\/}(1).

\bibitem[\protect\citeauthoryear{Dorfman, Hills, and Scharf}{Dorfman et~al.}{2022}]{dorfman2022guide}
Dorfman, A., T.~T. Hills, and I.~Scharf (2022).
\newblock A guide to area-restricted search: a foundational foraging behaviour.
\newblock {\em Biological Reviews\/}~{\em 97\/}(6), 2076--2089.

\bibitem[\protect\citeauthoryear{Drton and Plummer}{Drton and Plummer}{2017}]{drton_bayesian_2016}
Drton, M. and M.~Plummer (2017).
\newblock A bayesian information criterion for singular models.
\newblock {\em Journal of the Royal Statistical Society Series B: Statistical Methodology\/}~{\em 79\/}(2), 323--380.

\bibitem[\protect\citeauthoryear{Fan and Li}{Fan and Li}{2001}]{fan_variable_2001}
Fan, J. and R.~Li (2001).
\newblock Variable selection via nonconcave penalized likelihood and its oracle properties.
\newblock {\em Journal of the American Statistical Association\/}~{\em 96\/}(456), 1348--1360.

\bibitem[\protect\citeauthoryear{Florko, Togunov, Gryba, Sidrow, Ferguson, Yurkowski, and Auger-M{\'e}th{\'e}}{Florko et~al.}{2024}]{florko_review_nodate}
Florko, K., R.~R. Togunov, R.~Gryba, E.~Sidrow, S.~H. Ferguson, D.~J. Yurkowski, and M.~Auger-M{\'e}th{\'e} (2024).
\newblock A review of statistical models used to characterize species-habitat associations with animal movement data.
\newblock {\em arXiv preprint arXiv:2401.17389\/}.

\bibitem[\protect\citeauthoryear{Forney}{Forney}{1973}]{forney_viterbi_1973}
Forney, G. (1973).
\newblock The viterbi algorithm.
\newblock {\em Proceedings of the IEEE\/}~{\em 61\/}(3), 268--278.

\bibitem[\protect\citeauthoryear{Gassiat and Boucheron}{Gassiat and Boucheron}{2003}]{gassiat_optimal_2003}
Gassiat, E. and S.~Boucheron (2003).
\newblock Optimal error exponents in hidden markov models order estimation.
\newblock {\em IEEE Transactions on Information Theory\/}~{\em 49\/}(4), 964--980.

\bibitem[\protect\citeauthoryear{Glennie, Adam, Leos‐Barajas, Michelot, Photopoulou, and McClintock}{Glennie et~al.}{2023}]{glennie_hidden_2023}
Glennie, R., T.~Adam, V.~Leos‐Barajas, T.~Michelot, T.~Photopoulou, and B.~T. McClintock (2023).
\newblock Hidden {Markov} models: {Pitfalls} and opportunities in ecology.
\newblock {\em Methods in Ecology and Evolution\/}~{\em 14\/}(1), 43--56.

\bibitem[\protect\citeauthoryear{Hung, Wang, Zarnitsyna, Zhu, and Wu}{Hung et~al.}{2013}]{hung_hidden_2013}
Hung, Y., Y.~Wang, V.~Zarnitsyna, C.~Zhu, and C.~F.~J. Wu (2013).
\newblock Hidden {Markov} models with applications in cell adhesion experiments.
\newblock {\em Journal of the American Statistical Association\/}~{\em 108\/}(504), 1469--1479.

\bibitem[\protect\citeauthoryear{Hurford}{Hurford}{2009}]{hurford_gps_2009}
Hurford, A. (2009).
\newblock {GPS} {Measurement} {Error} {Gives} {Rise} to {Spurious} 180° {Turning} {Angles} and {Strong} {Directional} {Biases} in {Animal} {Movement} {Data}.
\newblock {\em PLoS ONE\/}~{\em 4\/}(5), e5632.

\bibitem[\protect\citeauthoryear{Kenyon, Yurkowski, Orr, Barber, and Ferguson}{Kenyon et~al.}{2018}]{kenyon_baffin_2018}
Kenyon, K.~A., D.~J. Yurkowski, J.~Orr, D.~Barber, and S.~H. Ferguson (2018).
\newblock Baffin {Bay} narwhal ({Monodon} monoceros) select bathymetry over sea ice during winter.
\newblock {\em Polar Biology\/}~{\em 41\/}(10), 2053--2063.

\bibitem[\protect\citeauthoryear{Leos-Barajas and Michelot}{Leos-Barajas and Michelot}{2018}]{leos-barajas_introduction_2018}
Leos-Barajas, V. and T.~Michelot (2018).
\newblock An introduction to animal movement modeling with hidden markov models using stan for bayesian inference.
\newblock {\em arXiv preprint arXiv:1806.10639\/}.

\bibitem[\protect\citeauthoryear{Li and Bolker}{Li and Bolker}{2017}]{li_incorporating_2017}
Li, M. and B.~M. Bolker (2017).
\newblock Incorporating periodic variability in hidden {Markov} models for animal movement.
\newblock {\em Movement Ecology\/}~{\em 5\/}(1), 1.

\bibitem[\protect\citeauthoryear{Lin and Song}{Lin and Song}{2022}]{lin_order_2022}
Lin, Y. and X.~Song (2022).
\newblock Order selection for regression-based hidden {Markov} model.
\newblock {\em Journal of Multivariate Analysis\/}~{\em 192}, 105061.

\bibitem[\protect\citeauthoryear{Mackay}{Mackay}{2002}]{mackay_estimating_2002}
Mackay, R.~J. (2002).
\newblock Estimating the order of a hidden markov model.
\newblock {\em Canadian Journal of Statistics\/}~{\em 30\/}(4), 573--589.

\bibitem[\protect\citeauthoryear{Manole and Khalili}{Manole and Khalili}{2021}]{manole_estimating_2021}
Manole, T. and A.~Khalili (2021).
\newblock Estimating the number of components in finite mixture models via the group-sort-fuse procedure.
\newblock {\em The Annals of Statistics\/}~{\em 49\/}(6), 3043--3069.

\bibitem[\protect\citeauthoryear{McClintock}{McClintock}{2021}]{mcclintock_worth_2021}
McClintock, B.~T. (2021).
\newblock Worth the effort? {A} practical examination of random effects in hidden {Markov} models for animal telemetry data.
\newblock {\em Methods in Ecology and Evolution\/}~{\em 12\/}(8), 1475--1497.

\bibitem[\protect\citeauthoryear{McClintock, Langrock, Gimenez, Cam, Borchers, Glennie, and Patterson}{McClintock et~al.}{2020}]{mcclintock_uncovering_2020}
McClintock, B.~T., R.~Langrock, O.~Gimenez, E.~Cam, D.~L. Borchers, R.~Glennie, and T.~A. Patterson (2020).
\newblock Uncovering ecological state dynamics with hidden {Markov} models.
\newblock {\em Ecology Letters\/}~{\em 23\/}(12), 1878--1903.

\bibitem[\protect\citeauthoryear{McClintock and Michelot}{McClintock and Michelot}{2018}]{mcclintock_momentuhmm_2018}
McClintock, B.~T. and T.~Michelot (2018).
\newblock {momentuHMM}: package for generalized hidden {Markov} models of animal movement.
\newblock {\em Methods in Ecology and Evolution\/}~{\em 9\/}(6), 1518--1530.

\bibitem[\protect\citeauthoryear{McKellar, Langrock, Walters, and Kesler}{McKellar et~al.}{2015}]{mckellar_using_2015}
McKellar, A.~E., R.~Langrock, J.~R. Walters, and D.~C. Kesler (2015).
\newblock Using mixed hidden {Markov} models to examine behavioral states in a cooperatively breeding bird.
\newblock {\em Behavioral Ecology\/}~{\em 26\/}(1), 148--157.

\bibitem[\protect\citeauthoryear{Morales, Haydon, Frair, Holsinger, and Fryxell}{Morales et~al.}{2004}]{morales_extracting_2004}
Morales, J.~M., D.~T. Haydon, J.~Frair, K.~E. Holsinger, and J.~M. Fryxell (2004).
\newblock {E}xtracting more out of relocation data: building movement models as mixtures of random walks.
\newblock {\em Ecology\/}~{\em 85\/}(9), 2436--2445.

\bibitem[\protect\citeauthoryear{Ngô, Heide-Jørgensen, and Ditlevsen}{Ngô et~al.}{2019}]{ngo_understanding_2019}
Ngô, M.~C., M.~P. Heide-Jørgensen, and S.~Ditlevsen (2019).
\newblock Understanding narwhal diving behaviour using {Hidden} {Markov} {Models} with dependent state distributions and long range dependence.
\newblock {\em PLOS Computational Biology\/}~{\em 15\/}(3), e1006425.

\bibitem[\protect\citeauthoryear{Patterson, McConnell, Fedak, Bravington, and Hindell}{Patterson et~al.}{2010}]{patterson_using_2010}
Patterson, T.~A., B.~J. McConnell, M.~A. Fedak, M.~V. Bravington, and M.~A. Hindell (2010).
\newblock Using {GPS} data to evaluate the accuracy of state–space methods for correction of {Argos} satellite telemetry error.
\newblock {\em Ecology\/}~{\em 91\/}(1), 273--285.

\bibitem[\protect\citeauthoryear{Pizzolato, Howell, Derksen, Dawson, and Copland}{Pizzolato et~al.}{2014}]{pizzolato_changing_2014}
Pizzolato, L., S.~E.~L. Howell, C.~Derksen, J.~Dawson, and L.~Copland (2014).
\newblock Changing sea ice conditions and marine transportation activity in {Canadian} {Arctic} waters between 1990 and 2012.
\newblock {\em Climatic Change\/}~{\em 123\/}(2), 161--173.

\bibitem[\protect\citeauthoryear{Pohle, Langrock, van Beest, and Schmidt}{Pohle et~al.}{2017}]{pohle_selecting_2017}
Pohle, J., R.~Langrock, F.~van Beest, and N.~M. Schmidt (2017).
\newblock Selecting the number of states in hidden markov models-pitfalls, practical challenges and pragmatic solutions.
\newblock {\em arXiv preprint arXiv:1701.08673\/}.

\bibitem[\protect\citeauthoryear{{R Core Team}}{{R Core Team}}{2021}]{R_2021}
{R Core Team} (2021).
\newblock {\em R: A Language and Environment for Statistical Computing}.
\newblock Vienna, Austria: R Foundation for Statistical Computing.

\bibitem[\protect\citeauthoryear{Schwarz}{Schwarz}{1978}]{schwarz_estimating_1978}
Schwarz, G. (1978).
\newblock Estimating the dimension of a model.
\newblock {\em The Annals of Statistics\/}~{\em 6\/}(2).

\bibitem[\protect\citeauthoryear{Shuert, Hussey, Marcoux, Heide-Jørgensen, Dietz, and Auger-Méthé}{Shuert et~al.}{2023}]{shuert_divergent_2023}
Shuert, C.~R., N.~E. Hussey, M.~Marcoux, M.~P. Heide-Jørgensen, R.~Dietz, and M.~Auger-Méthé (2023).
\newblock Divergent migration routes reveal contrasting energy-minimization strategies to deal with differing resource predictability.
\newblock {\em Movement Ecology\/}~{\em 11\/}(1), 31.

\bibitem[\protect\citeauthoryear{Shuert, Marcoux, Hussey, Heide-Jørgensen, Dietz, and Auger-Méthé}{Shuert et~al.}{2022}]{shuert_decadal_2022}
Shuert, C.~R., M.~Marcoux, N.~E. Hussey, M.~P. Heide-Jørgensen, R.~Dietz, and M.~Auger-Méthé (2022).
\newblock Decadal migration phenology of a long-lived {Arctic} icon keeps pace with climate change.
\newblock {\em Proceedings of the National Academy of Sciences\/}~{\em 119\/}(45), e2121092119.

\bibitem[\protect\citeauthoryear{Smyth}{Smyth}{2000}]{smyth_model_2000}
Smyth, P. (2000).
\newblock Model selection for probabilistic clustering using cross-validated likelihood.
\newblock {\em Statistics and computing\/}~{\em 10\/}(1), 63--72.

\bibitem[\protect\citeauthoryear{Sutherland}{Sutherland}{1998}]{sutherland_importance_1998}
Sutherland, W.~J. (1998).
\newblock The importance of behavioural studies in conservation biology.
\newblock {\em Animal Behaviour\/}~{\em 56\/}(4), 801--809.

\bibitem[\protect\citeauthoryear{Viterbi}{Viterbi}{1967}]{viterbi_error_1967}
Viterbi, A. (1967).
\newblock Error bounds for convolutional codes and an asymptotically optimum decoding algorithm.
\newblock {\em IEEE Transactions on Information Theory\/}~{\em 13\/}(2), 260--269.

\bibitem[\protect\citeauthoryear{Wang, Li, and Tsai}{Wang et~al.}{2007}]{wang_tuning_2007}
Wang, H., R.~Li, and C.-L. Tsai (2007).
\newblock Tuning parameter selectors for the smoothly clipped absolute deviation method.
\newblock {\em Biometrika\/}~{\em 94\/}(3), 553--568.

\bibitem[\protect\citeauthoryear{Watanabe}{Watanabe}{2013}]{watanabe_widely_2013}
Watanabe, S. (2013).
\newblock A widely applicable bayesian information criterion.
\newblock {\em The Journal of Machine Learning Research\/}~{\em 14\/}(1), 867--897.

\bibitem[\protect\citeauthoryear{Watt and Ferguson}{Watt and Ferguson}{2015}]{watt_fatty_2015}
Watt, C.~A. and S.~H. Ferguson (2015).
\newblock Fatty acids and stable isotopes ($\delta$ $^{\textrm{13}}$ {C} and $\delta$ $^{\textrm{15}}$ {N}) reveal temporal changes in narwhal ( \textit{{Monodon} monoceros} ) diet linked to migration patterns.
\newblock {\em Marine Mammal Science\/}~{\em 31\/}(1), 21--44.

\bibitem[\protect\citeauthoryear{Whoriskey, Auger‐Méthé, Albertsen, Whoriskey, Binder, Krueger, and Mills~Flemming}{Whoriskey et~al.}{2017}]{whoriskey_hidden_2017}
Whoriskey, K., M.~Auger‐Méthé, C.~M. Albertsen, F.~G. Whoriskey, T.~R. Binder, C.~C. Krueger, and J.~Mills~Flemming (2017).
\newblock A hidden {Markov} movement model for rapidly identifying behavioral states from animal tracks.
\newblock {\em Ecology and Evolution\/}~{\em 7\/}(7), 2112--2121.

\bibitem[\protect\citeauthoryear{Zou and Li}{Zou and Li}{2008}]{zou_one-step_2008}
Zou, H. and R.~Li (2008).
\newblock One-step sparse estimates in nonconcave penalized likelihood models.
\newblock {\em The Annals of Statistics\/}~{\em 36\/}(4).

\bibitem[\protect\citeauthoryear{Zucchini, MacDonald, and Langrock}{Zucchini et~al.}{2017}]{zucchini_hidden_2017}
Zucchini, W., I.~L. MacDonald, and R.~Langrock (2017).
\newblock {\em Hidden Markov models for time series: an introduction using R}.
\newblock CRC press.

\end{thebibliography}
\newpage

\renewcommand\thefigure{A\arabic{figure}} 
\renewcommand\thetable{A\arabic{table}} 
\setcounter{figure}{0} 
\setcounter{table}{0} 
\begin{appendices}
\section{Supporting Information}
\subsection{Procedure to get tpm estimates}
To obtain an estimate of the state-transition probabilities from the DPMLE, the average of the state-transition probabilities can be used.
\label{Two}
Let $\widehat{\Gamma}$ be the estimated tpm obtained from the double penalized log-likelihood method such that:
$$
\widehat{\Gamma}=\begin{pmatrix}
    \hat{\gamma}_{11}&\hat{\gamma}_{12}&\hat{\gamma}_{13}&\hat{\gamma}_{14}\\
    \hat{\gamma}_{21}&\hat{\gamma}_{22}&\hat{\gamma}_{23}&\hat{\gamma}_{24}\\
    \hat{\gamma}_{31}&\hat{\gamma}_{32}&\hat{\gamma}_{33}&\hat{\gamma}_{34}\\
    \hat{\gamma}_{41}&\hat{\gamma}_{42}&\hat{\gamma}_{43}&\hat{\gamma}_{44}
    
\end{pmatrix}
$$
with $\hat{\mu}_1=\hat{\mu}_2$ and $\hat{\mu}_3=\hat{\mu}_4$ such that $s_1$ (respectively $s_3$) and $s_2$ (respectively $s_4$) are merged into one state $\tilde{s}_1$ (respectively $\tilde{s}_2$).

An estimate for the final probability $\tilde{\gamma}_{11}$ is therefore:

\begin{center}
$
\begin{array}{lll}
    \tilde{\gamma}_{11} &=&  \mathbb{P}[\tilde{s}_1 \to \tilde{s}_1]\\
    
    &=&  (\mathbb{P}[s_1 \to s_1 \text{ or } s_1 \to s_2] + \mathbb{P}[s_2 \to s_2 \text{ or } s_2 \to s_1])/2 \\
     &=&(\hat{\gamma}_{11}+\hat{\gamma}_{12}+\hat{\gamma}_{21}+\hat{\gamma}_{22})/2.
\end{array}
$
\end{center}
The same method can be used to obtain $\tilde{\gamma}_{22}$ and thus $\widetilde{\Gamma}.$

\subsection{Viterbi algorithm to estimate $\hat{\boldsymbol{\pi}}(\boldsymbol{\Psi})$}
\label{viterbistationary}
To derive $\hat{\pi}_j(\boldsymbol{\Psi})$, the Viterbi algorithm can be used to compute the following:
\begin{center}
\begin{equation}
\label{ViterbiEstimate}
\hat{\pi}_j(\boldsymbol{\Psi})=\widehat{V}_{j}(T,\boldsymbol{\Psi}),
\end{equation}
\end{center}
$\widehat{V}_{j}(T,\boldsymbol{\Psi}) = \frac{\#\{t\leq T, S_{t}=j\}}{T}= \frac1T\underset{t=1}{\overset{T}\sum}\mathbbm{1}_{S_t=j}$. It can be estimated using the most likely state sequence returned by the Viterbi algorithm, which depends on the observations and parameter estimates (\citealp{viterbi_error_1967}; \citealp{forney_viterbi_1973}). In this case, the gradient of $\hat{\pi}_j(\boldsymbol{\hat{\delta}}^{(p)},\boldsymbol{\beta},\boldsymbol{\hat{\mu}}^{(p)},\boldsymbol{\hat{\sigma}}^{(p)})$ with respect to $\boldsymbol{\beta}$ is challenging to compute therefore the function \texttt{optim} should be used, with the default Nelder-Mead optimization method that does not require the gradient. The Nelder-Mead approach is slower than gradient-based methods, and thus the computational time of the procedure is higher in the non-stationary case. 
\subsection{Computational complexity of the gradient of $\hat{\pi}_j(\boldsymbol{\Psi})$ estimated with forward-backward probabilities.}
 The non-stationary DPMLE is more flexible but involves more operations to update the transition matrix than in the stationary DPMLE. We compare both computational costs.
 Consider Eq (\ref{tildepen}) and (\ref{tilderpennh}) to update the transition matrix for the stationary (without covariates) and non-stationary (with time-varying covariates) DPMLE as follows:

\begin{center}
\begin{equation}
\label{tildepen}
   {\text{argmax}}_{\boldsymbol{\Gamma}}\underset{j=1}{\overset{N}{\sum}}\hat{u}_{j}^{(p)}(1)\log{\pi_{j}}
    + C_N\underset{j=1}{\overset{N}{\sum}}\log{\pi_{j}}
    +  \underset{t=2}{\overset{T}{\sum}}
    \underset{j,i=1}{\overset{N}{\sum}}\hat{v}_{ij}^{(p)}(t)\log{\gamma_{ij}},
\end{equation}
\end{center}
and 
\begin{center}
\begin{equation}
\label{tilderpennh}
   {\text{argmax}}_{\boldsymbol{\beta}}
    C_N\underset{j=1}{\overset{N}{\sum}}\log{\hat{\pi}_{j}}
    +  \underset{t=2}{\overset{T}{\sum}}
    \underset{j,i=1}{\overset{N}{\sum}}\hat{v}_{ij}^{(p)}(t)\log{\gamma_{ij}^{(t)}}.
\end{equation}
\end{center}
with ${\boldsymbol{\beta}}$ defined by Eq (\ref{t.p.m.t}) and Eq (\ref{t.p.m.tt}).

Consider first the right-hand side of both equations. In Eq (\ref{tildepen}), each element of $\boldsymbol{\Gamma}$ consists of dividing one element $ e^{c_{ij}}$ by the sum of $N-1$ elements $\underset{k\neq i}{ \overset{N}{\sum}} e^{c_{ik}}$ which is of computational complexity $O(N-1)= O(N)$, and the number of operations involved to compute $\boldsymbol{\Gamma}$ is of order $O(N^3)$.

Thus with $C$ time-varying covariates, it requires $O(N^3C)$ operations to compute $\boldsymbol{\Gamma}_t$. The factor $C$ is due to the fact that in Eq (\ref{t.p.m.t}), each element $c_{\underset{i\neq j}{ij}}^{(t)}=
\beta_0^{ij}+\underset{c=1}{\overset{C}{\sum}}\beta_c^{ij}\omega_c^{(t)}$
is the sum of $C+1$ elements. Thus there are $O(TCN^3)$ operations involved to compute 
 $(\boldsymbol{\Gamma}^{(t)})_{t={1,\ldots,T}}.$ We examine the scenario in which the stationary DPMLE has no covariates, while the non-stationary DPMLE includes time-varying covariates. This is because in the case of the stationary method with covariates, they would also be included in the non-stationary approach (e.g., covariates such as sex and age category). Consequently, focusing exclusively on the case without constant covariates is sufficient to compare the difference in computational cost between both methods.

In the stationary DPMLE $\boldsymbol{\pi}$ arises from $\boldsymbol{\Gamma}$ as follows:
$$\boldsymbol{\pi}= \boldsymbol{1}(I_{N}-\boldsymbol{\Gamma}+\boldsymbol{U})^{-1},$$ which has computational complexity of order $N^3$. In the non-stationary DPMLE, $\boldsymbol{\pi}$ is estimated with the forward and backward probabilities as follows:
$$\hat{\pi}_j(\boldsymbol{\Psi}) = \frac{1}{T}\underset{t=1}{\overset{T}{\sum}}\mathbb{P}[S_t=j|\boldsymbol{Y}=\boldsymbol{y}]= \frac{1}{TL}\underset{t=1}{\overset{T}{\sum}}\alpha_{t}(j)\beta_{t}(j).$$

Each element of $\hat{\pi}(\boldsymbol{\Psi})$ is the sum of $T$ elements. ${\alpha}_t(j)$ is a sum of $N$ products of three quantities: an element of $\boldsymbol{\alpha}_{t-1}$, a transition probability $\gamma_{ij}^{(t)}$ (already computed), and a state-dependent probability $f_j(\boldsymbol{Y}_t=\boldsymbol{y}_t|S_t=j;\theta_j)$. 

Similarly for $\beta_t,$ and $L.$ Thus $O(N^2T)$ operations are requires to compute $\hat{\pi}(\boldsymbol{\Psi}).$ Thus if the gradient is computed with finite differences, the non-stationary DPMLE has computational complexity $O(N^5C^2T)$ and the stationary DPMLE $O(N^5)$.
\subsection{Additional results}
\begin{figure}[H]
 \begin{subfigure}[b]{0.5\textwidth}  
     \centering
    \includegraphics[width=0.5\textwidth]{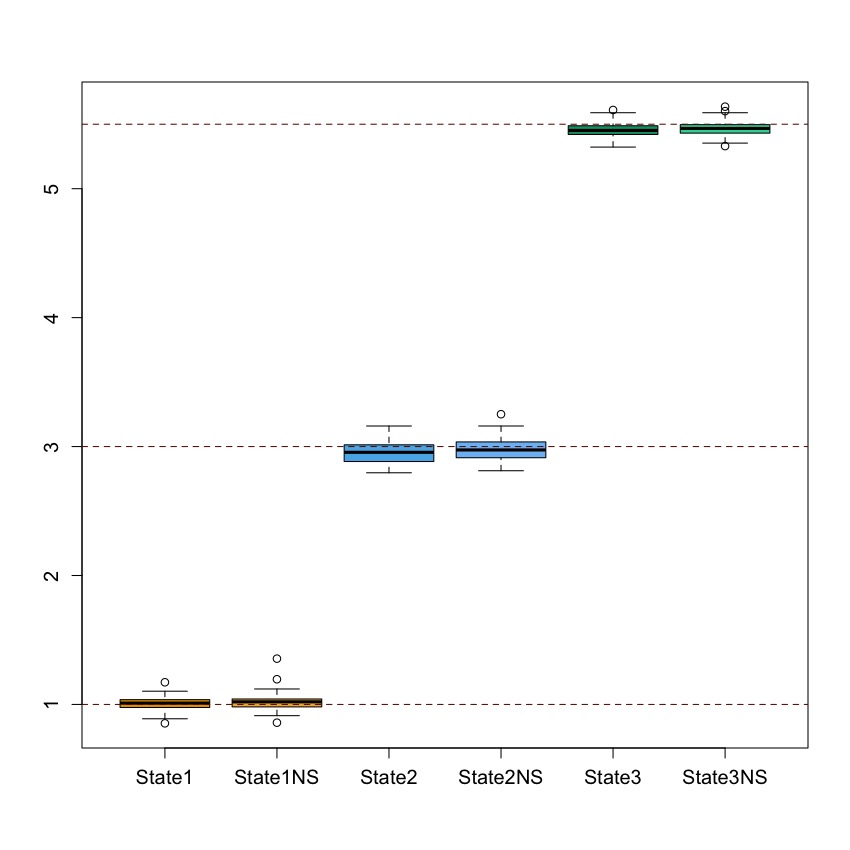}
        \label{}
        \caption{}
    \end{subfigure}
      \begin{subfigure}[b]{0.5\textwidth}
          \centering
\includegraphics[width=0.5\textwidth]{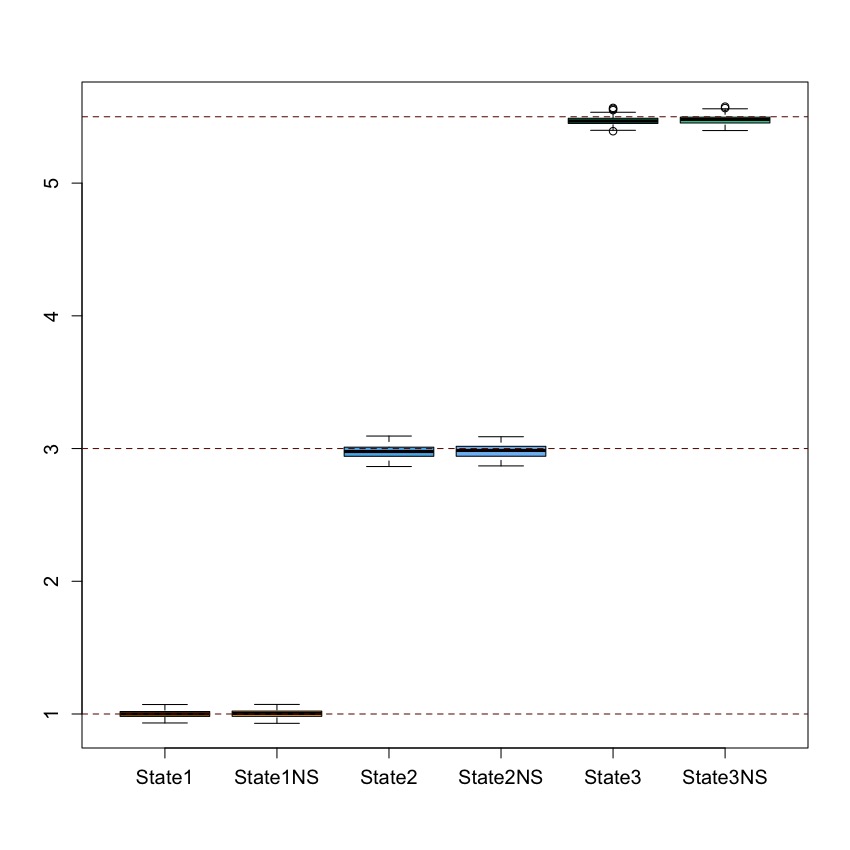}
\caption{}
        \label{}
    \end{subfigure}
    \vskip1cm
       \begin{subfigure}[b]{0.5\textwidth}
                  \centering
\includegraphics[width=0.5\textwidth]{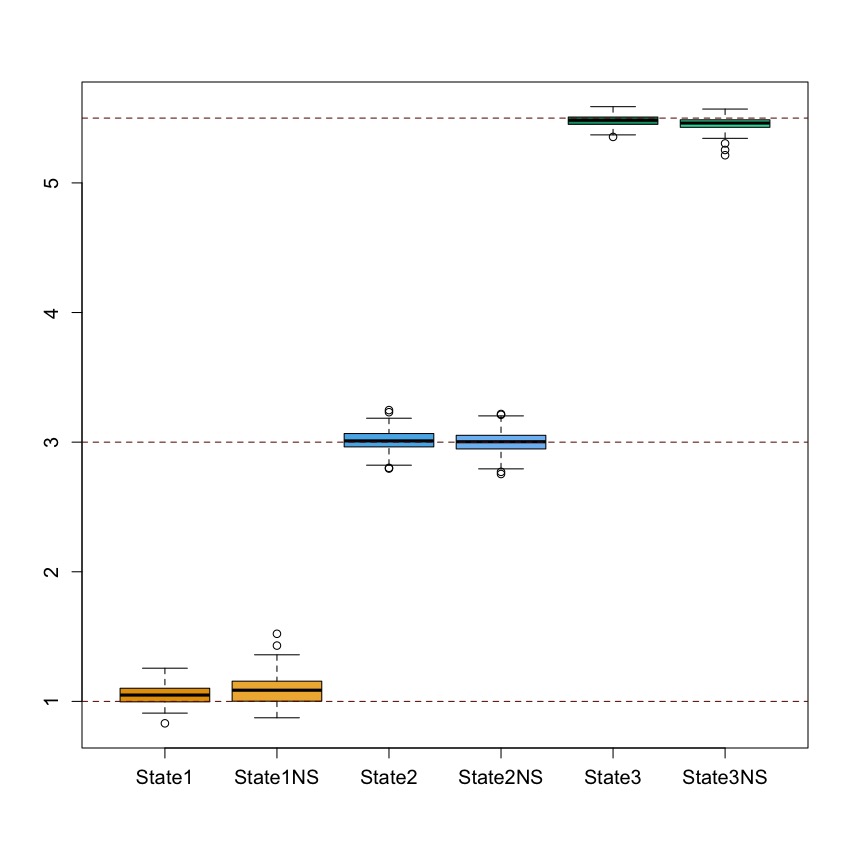}
        \caption{}
    \end{subfigure}
          \begin{subfigure}[b]{0.5\textwidth}
          \centering
\includegraphics[width=0.5\textwidth]{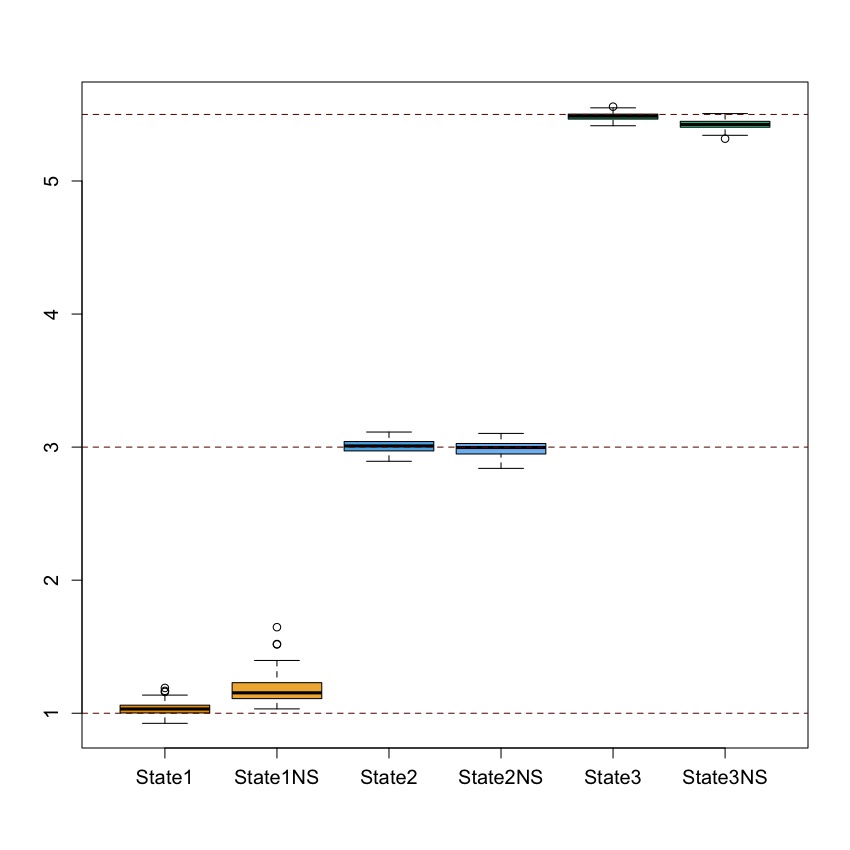}
\caption{}
        \label{}
    \end{subfigure}
        \vskip1cm

       \begin{subfigure}[b]{0.5\textwidth}
                  \centering
\includegraphics[width=0.5\textwidth]{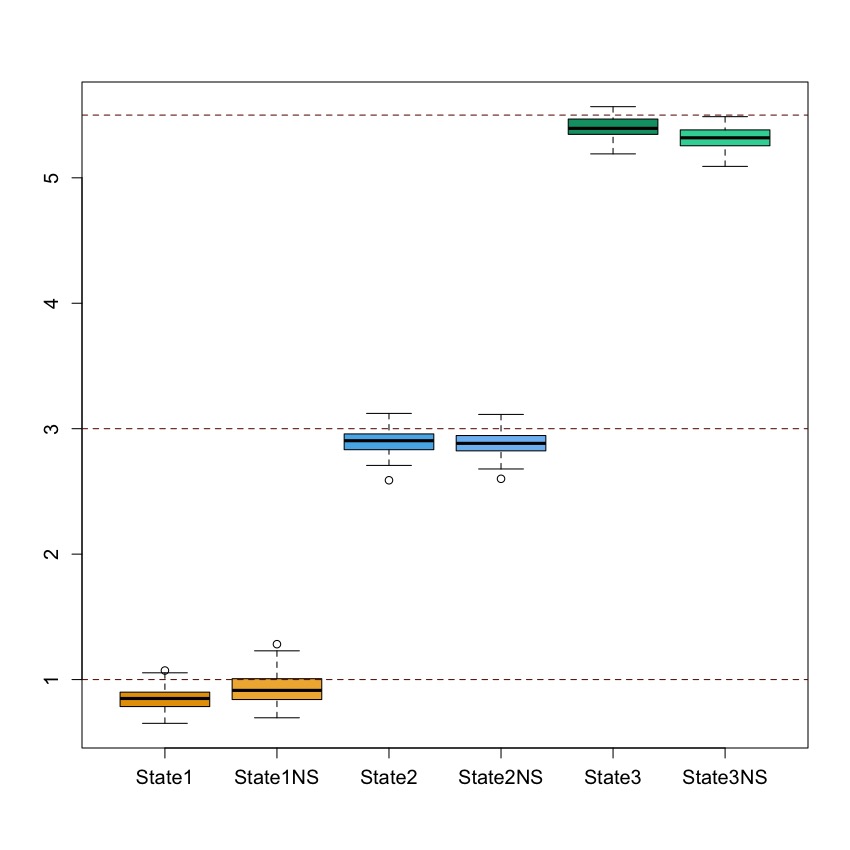}
        \caption{}
    \end{subfigure}
       \begin{subfigure}[b]{0.5\textwidth}
                  \centering
\includegraphics[width=0.5\textwidth]{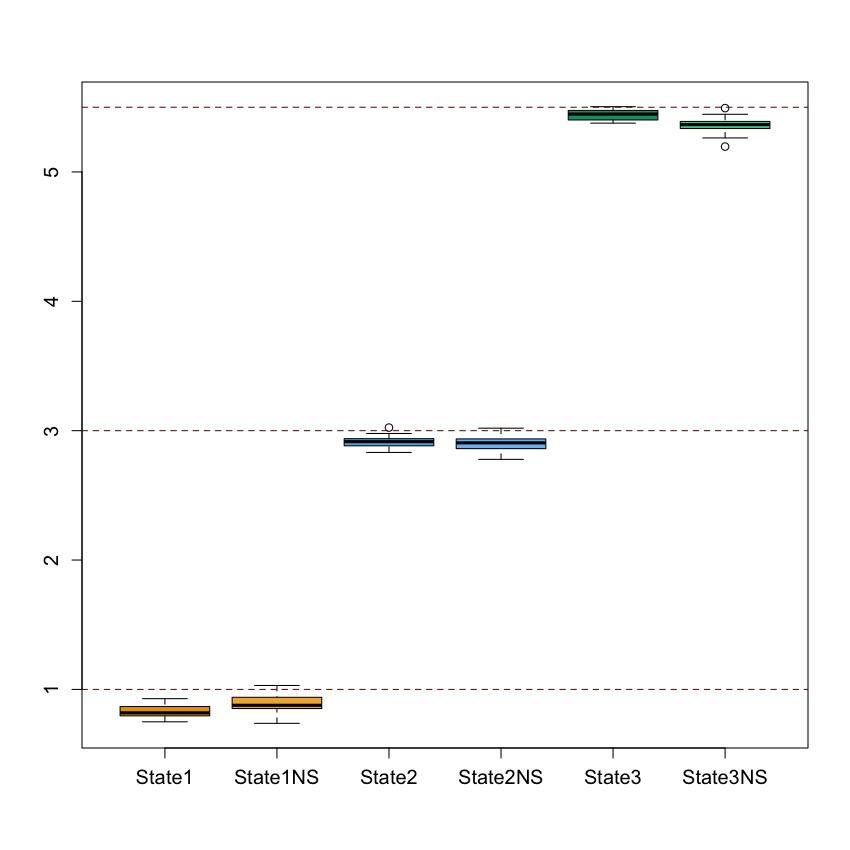}
        \caption{}
    \end{subfigure}
        \vskip1cm

    \begin{subfigure}[b]{0.5\textwidth}
                  \centering
\includegraphics[width=0.5\textwidth]{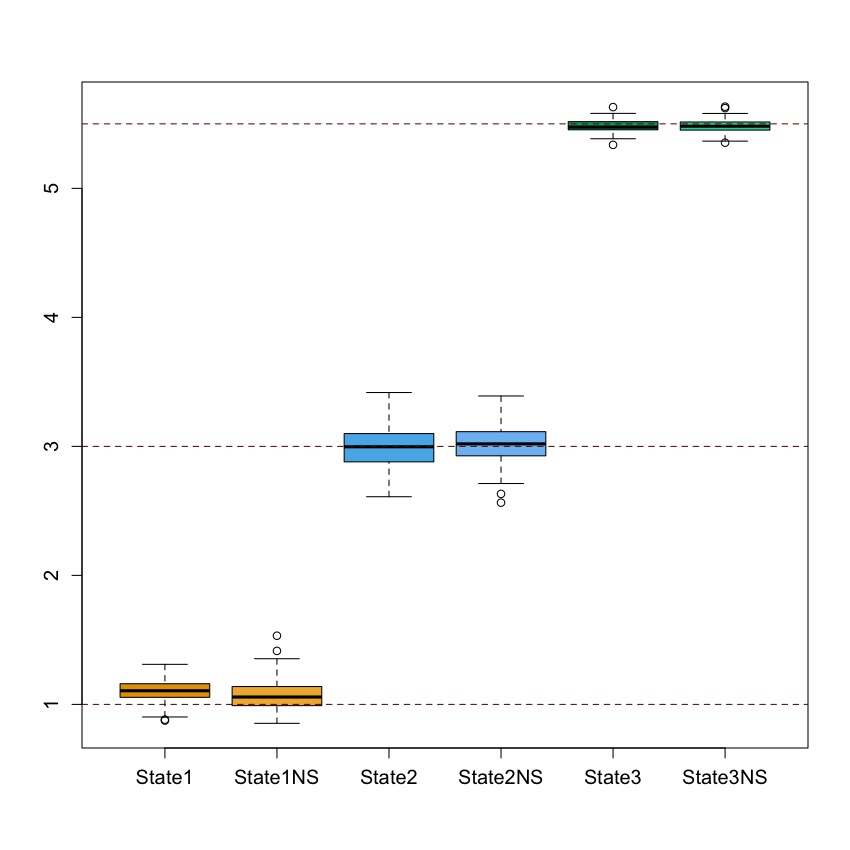}
        \caption{}
    \end{subfigure}
       \begin{subfigure}[b]{0.5\textwidth}
                  \centering
\includegraphics[width=0.5\textwidth]{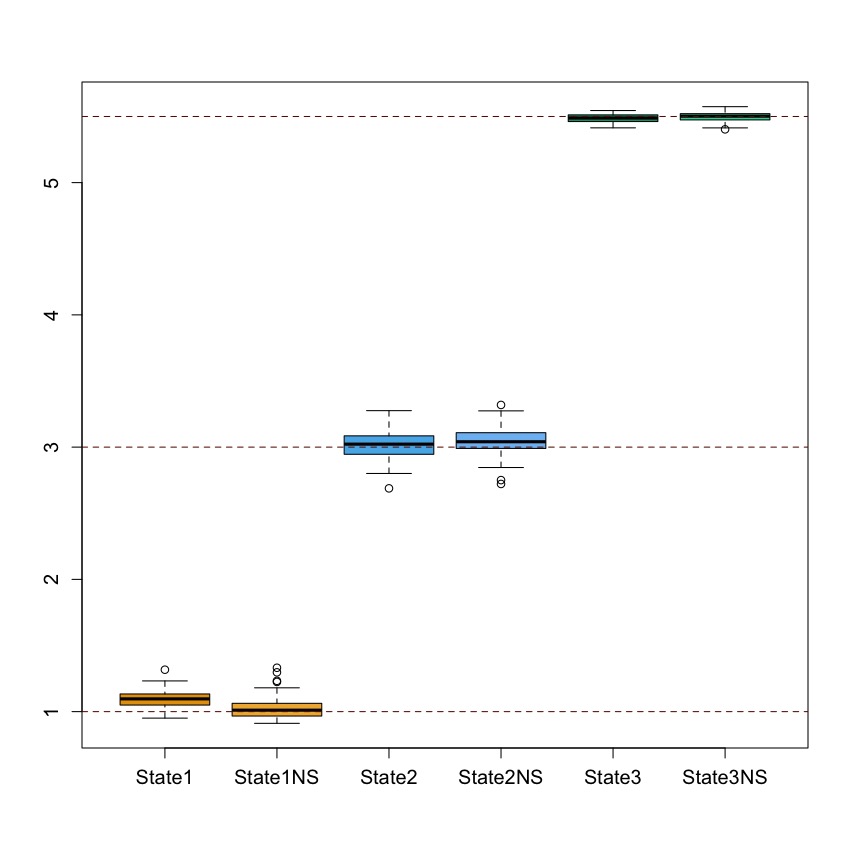}
        \caption{}
    \end{subfigure}
    \caption{\label{boxplots}Boxplots of the $100$ mean estimates obtained by the DPMLEs methods for each state. Data were generated under scenario 2 (outliers; top), 3 (heterogeneity in tpm; middle), 5 (autocorrelation in emission; bottom middle) and 6 (temporal variation in hidden process; bottom) with $5000$ (left boxplot) and $12000$ (right boxplot) observations. "stateNS" refers to the non-stationary DPMLE.}
    \end{figure}
\begin{figure}[H]
 \begin{subfigure}[b]{0.5\textwidth}  
     \centering
    \includegraphics[width=0.5\textwidth]{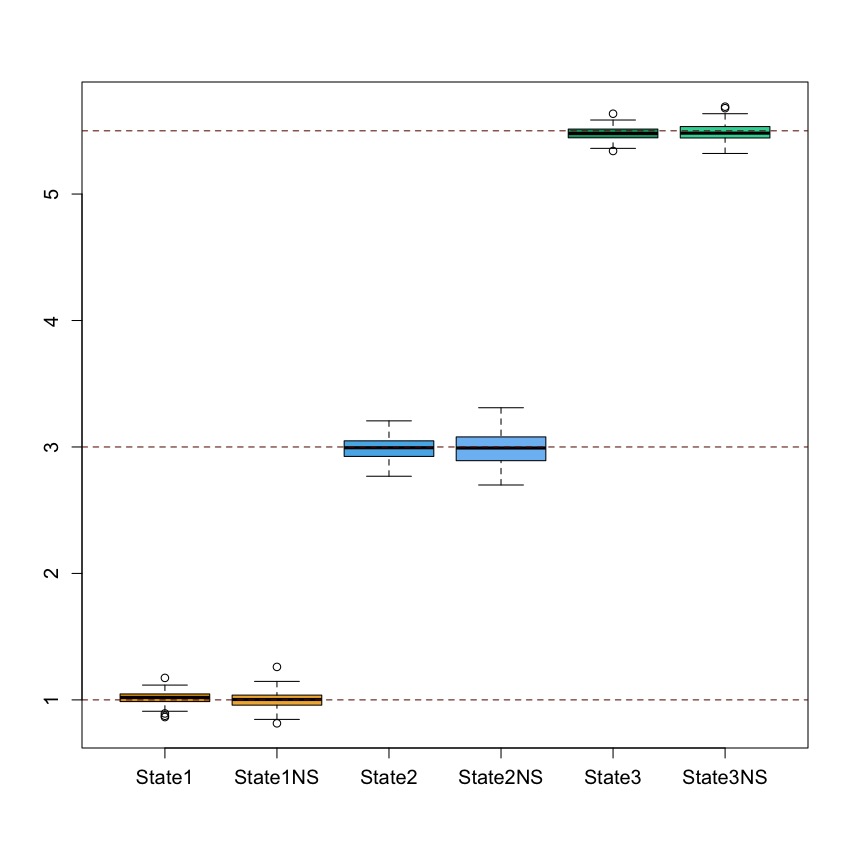}
        \label{}
        \caption{}
    \end{subfigure}
      \begin{subfigure}[b]{0.5\textwidth}
          \centering
\includegraphics[width=0.5\textwidth]{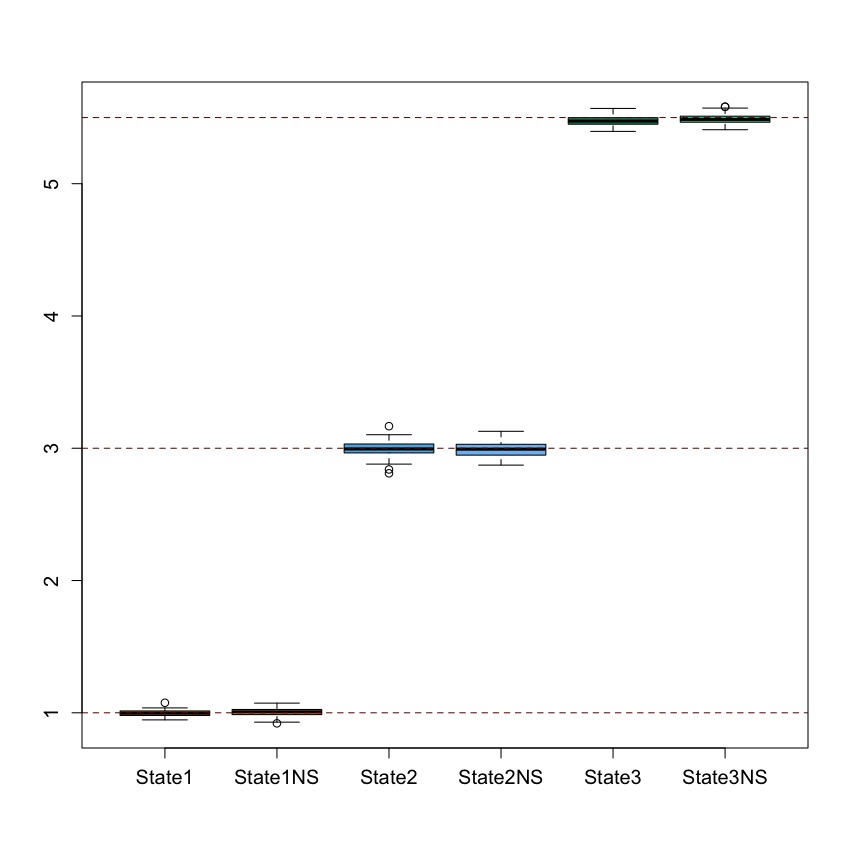}
        \label{}
                \caption{}

    \end{subfigure}
        \caption{\label{boxplotsbenchmark}Boxplots of the $100$ mean estimates obtained by the DPMLEs methods for each state. Data were generated under scenario 1 (benchmark) with $T=5,000$ (left boxplot) and $T=12,000$ (right boxplot)."stateNS" refers to the non-stationary DPMLE. }
\end{figure}

\subsection{Narwhal data processing}
\label{dataprocess}
\begin{figure}[H]
    \centering
    \includegraphics[width = 15cm, height = 12cm]{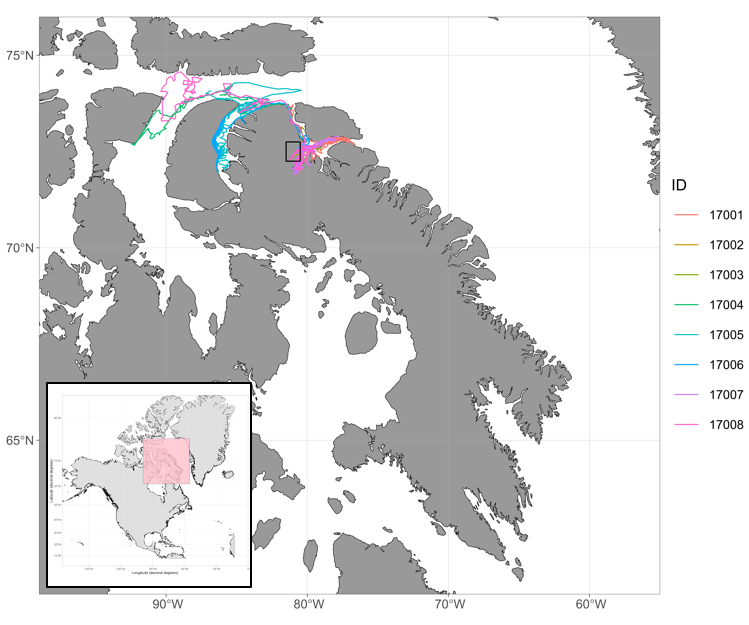}
    \caption{Map of the location
    data of the eight individuals from August 2017 to September 2017, after data cleaning. Each colour corresponds to an individual. The black box is Tremblay Sound, where the narwhal were tagged.}
    \label{fig:narwhal1}
\end{figure}

We split tracks with gaps larger than $12$ hours. More specifically, we set an individual ID when there were more than $12$ consecutive missing data points and the proportion of missing data fell below $50\%$. The underlying assumption is that states are independent of one another when they are separated by more than $12$ hours. We filtered out segments that were too short to properly fit an HMM by using a threshold of $6$ locations. Since the original resolution was finer at some locations, we rounded the values to the closest hour. If more than one location was attributed to the same time point, we picked the last one (i.e. most recent) as the final value. This process created $18$ "individuals", some of them representing the movement of the same narwhal, but with a significant time interval (i.e., at least $12$ hours apart) between them (a map of the tracks before data cleaning can be found in Fig$.$ \ref{fig:narwhal1}).

\subsection{Narwhal case study: additional results}
\begin{figure}[H]
 \begin{subfigure}[b]{0.5\textwidth}  
     \centering
    \includegraphics[width=0.75\textwidth]{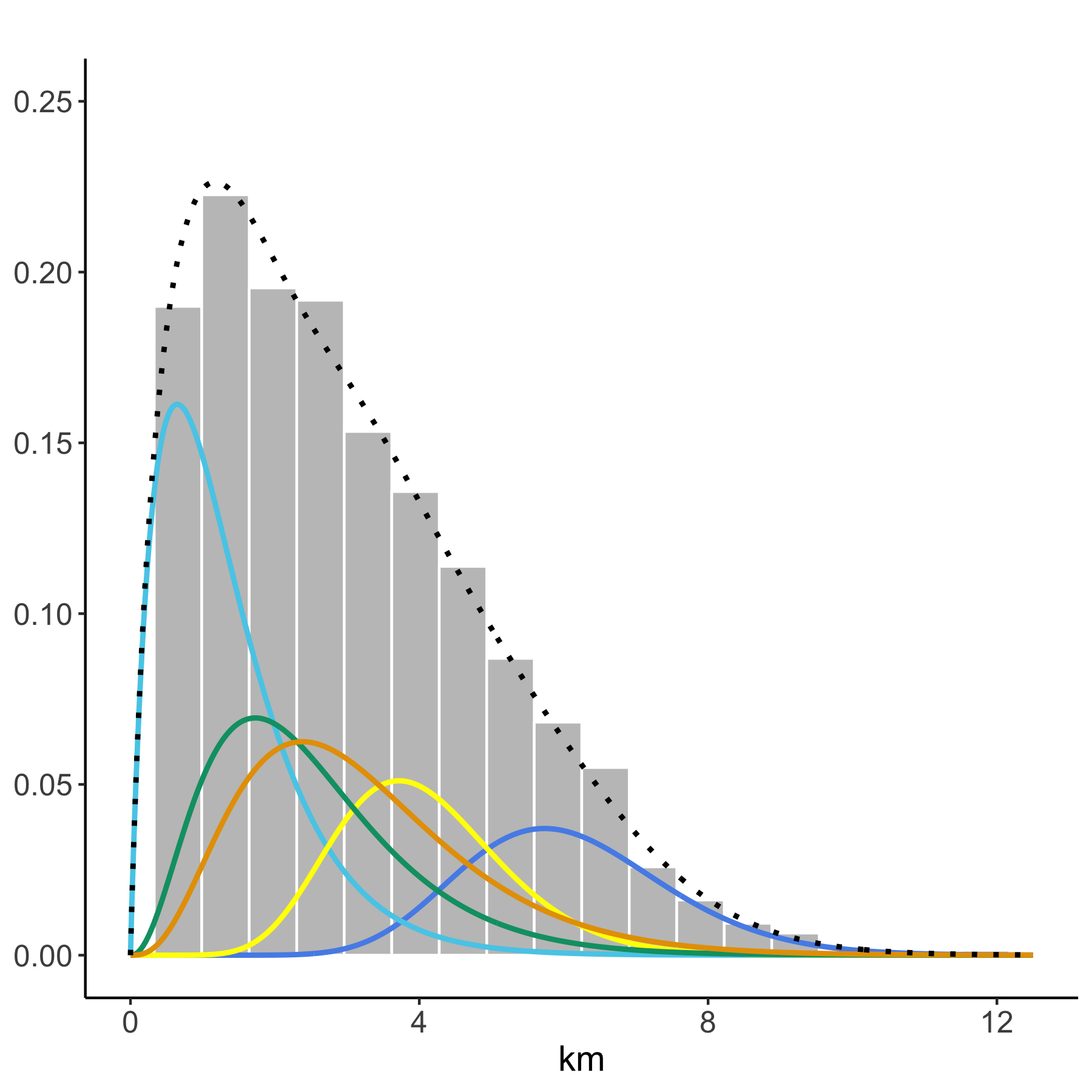}
        \label{}
    \end{subfigure}
      \begin{subfigure}[b]{0.5\textwidth}
          \centering

\includegraphics[width=0.75\textwidth]{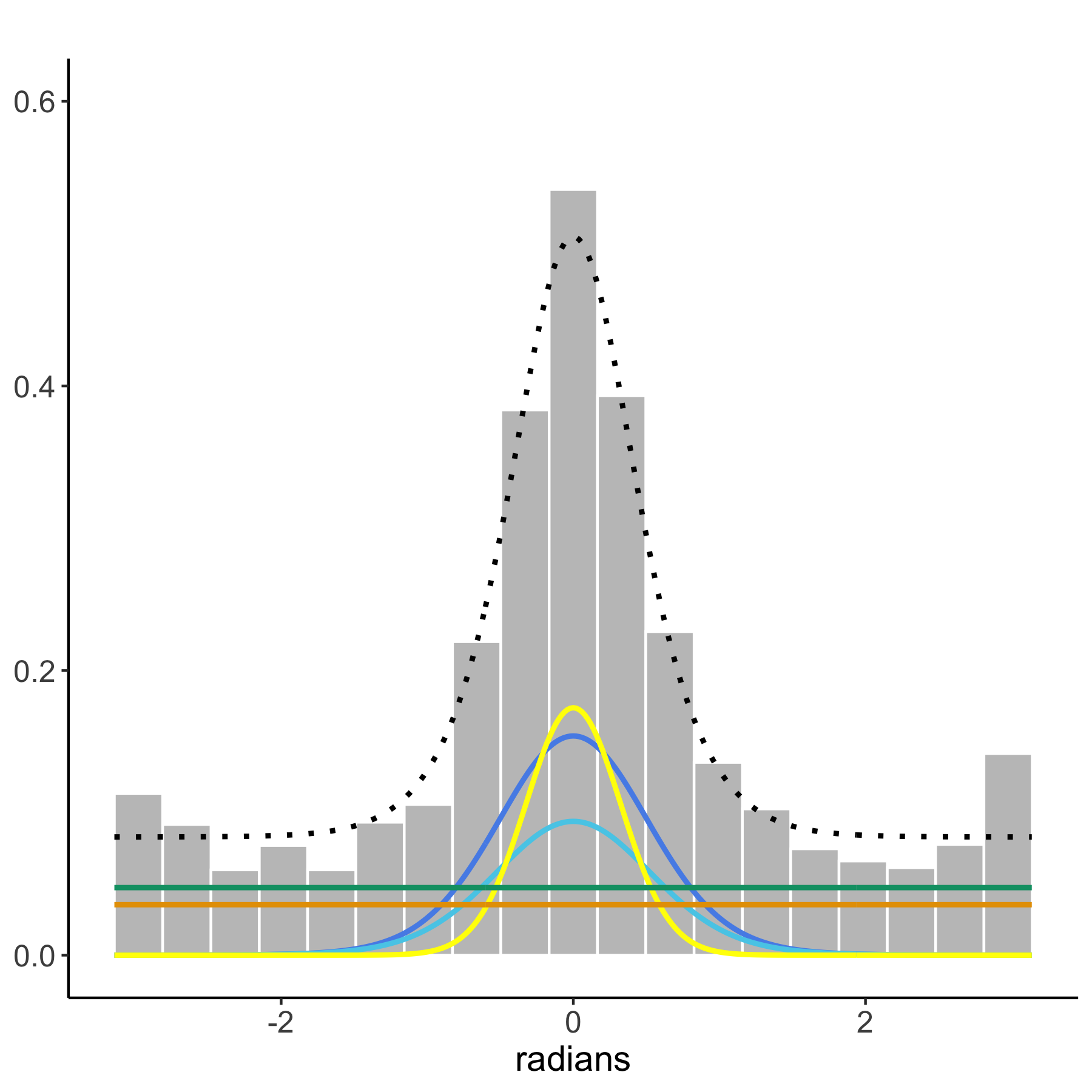}
        \label{}
    \end{subfigure}
\caption{\label{BICfit}Histograms of the step length and turning angles. Curves represent the estimated distributions from the model selected by BIC. Each colour corresponds to a different state.}
\end{figure}
\subsection{Additional Results}

\begin{table}[H]
\small
\begin{center}
\begin{tabular}{|c|l|c |c|c|}
\specialrule{.3em}{.2em}{.2em}
\multicolumn{5}{c}{True Number of States = 3}\\
\specialrule{.3em}{.2em}{.2em}
  Scenario &Criterion &\multicolumn{3}{l}{Number of states selected} \\[0.05cm]
\hline
\multirow{6}{*}{1: Benchmark}
&& 2 (\%)&3 (\%)&4 (\%)\\[0.2cm]
 &AIC&0&16&\textbf{84}\\
  &AIC cov&0&30&\textbf{70}\\
 &BIC &0&\textbf{100}&0\\
  &BIC cov&0&\textbf{100}&0\\
  &DPMLE&0&\textbf{100}&0\\
  &DPMLE cov&0&\textbf{100}&0\\
\hline

\multirow{6}{*}{2: outliers}
 &AIC&0&0&\textbf{100}\\
 &AIC cov& 0&0&\textbf{100}\\
 &BIC&0&0&\textbf{100}\\
 &BIC cov&0&0&\textbf{100}\\
  &DPMLE&0&\textbf{100}&0\\
  &DPMLE cov&0&\textbf{100}&0\\
\hline

\multirow{6}{*}{3: heterogeneity in tpm}
 &AIC&0&0&\textbf{100}\\
 &AIC cov& 0&0&\textbf{100}\\
 &BIC&0&16&\textbf{84}\\
 &BIC cov&0&\textbf{52}&48\\
  &DPMLE&0&\textbf{99}&1\\
  &DPMLE cov&0&\textbf{100}&0\\
\hline

\multirow{6}{*}{4: heterogeneity in emission}
 &AIC&0&0&\textbf{100}\\
 &AIC cov& 0&21&\textbf{79}\\
 &BIC&0&\textbf{55}&45\\
 &BIC cov&0&\textbf{87}&13\\
  &DPMLE&0&\textbf{69}&31\\
  &DPMLE cov&0&\textbf{71}&29\\
\hline

\multirow{6}{*}{5: violation of conditional independence}
 &AIC&0&0&\textbf{100}\\
 &AIC cov& 0&0&\textbf{100}\\
 &BIC&0&1&\textbf{99}\\
 &BIC cov&0&\textbf{100}&0\\
  &DPMLE&0&\textbf{76}&24\\
  &DPMLE cov&0&\textbf{99}&1\\
\end{tabular}
\end{center}
\caption*{}
\end{table}

\begin{table}[H]
\small
\begin{center}
\begin{tabular}{|c|l|c |c|c|}
\textcolor{white}{5: violation of conditional independence}&\textcolor{white}{Criterion}& \textcolor{white}{2 (\%)}&\textcolor{white}{3 (\%)}&\textcolor{white}{ 4 (\%)}\\
\multirow{5}{*}{6: temporal variation}
 &AIC&0&11&\textbf{89}\\
 &AIC cov& 0&0&\textbf{100}\\
 &BIC&0&\textbf{100}&0\\
 &BIC cov&0&\textbf{97}&3\\
  &DPMLE&0&\textbf{100}&0\\
  &DPMLE cov&0&\textbf{100}&0 \\ \specialrule{.3em}{.2em}{0em}
\end{tabular}
\end{center}

\caption{\label{allresultsthree}
Percentages of the $100$ simulation trials in 2-4 number of hidden states that were chosen by different criteria: AIC, BIC and DPMLE. "cov" refers to the models fitted with the linear covariate "time of day" in the transition probabilities. Every track was simulated with $5,000$ observations according to the simulation scenario specified.}
\end{table}

\begin{table}[H]
\small
\begin{center}
\begin{tabular}{|c|l|c |c|c|}
\specialrule{.3em}{.2em}{.2em}
\multicolumn{5}{c}{True Number of States = 3}\\
\specialrule{.3em}{.2em}{.2em}
  Scenario &Criterion &\multicolumn{3}{l}{Number of states selected} \\[0.05cm]
  \hline

\multirow{6}{*}{1: Benchmark}
 &AIC&0&21&\textbf{79}\\
 &AIC cov& 0&\textbf{79}&21\\
 &BIC&0&\textbf{100}&0\\
 &BIC cov&0&\textbf{100}&0\\
  &DPMLE&0&\textbf{100}&0\\
  &DPMLE cov&0&\textbf{100}&0\\
\hline
\multirow{6}{*}{2: outliers}
 &AIC&0&0&\textbf{100}\\
 &AIC cov& 0&0&\textbf{100}\\
 &BIC&0&0&\textbf{100}\\
 &BIC cov&0&0&\textbf{100}\\
  &DPMLE&0&\textbf{100}&0\\
  &DPMLE cov&0&\textbf{100}&0\\
\hline

\multirow{6}{*}{3: heterogeneity in tpm}
 &AIC&0&0&\textbf{100}\\
 &AIC cov& 0&0&\textbf{100}\\
 &BIC&0&2&\textbf{98}\\
 &BIC cov&0&11&\textbf{89}\\
  &DPMLE&0&\textbf{100}&0\\
  &DPMLE cov&0&\textbf{100}&0\\
\hline

\multirow{6}{*}{4: heterogeneity in emission}
 &AIC&0&0&\textbf{100}\\
 &AIC cov& 0&10&\textbf{90}\\
 &BIC&0&20&\textbf{80}\\
 &BIC cov&0&43&\textbf{57}\\
  &DPMLE&0&{23}&\textbf{77}\\
  &DPMLE cov&0&\textbf{67}&33\\
\hline

\multirow{6}{*}{5: violation of conditional independence}
 &AIC&0&0&\textbf{100}\\
 &AICcov& 0&0&\textbf{100}\\
 &BIC&0&0&\textbf{100}\\
 &BICcov&0&47&\textbf{53}\\
  &DPMLE&0&21&\textbf{79}\\
  &DPMLE cov&0&\textbf{85}&15\\
\hline

\multirow{6}{*}{6: temporal variation in tpm}
 &AIC&0&6&\textbf{94}\\
 &AIC cov& 0&0&\textbf{100}\\
 &BIC&0&\textbf{100}&0\\
 &BIC cov&0&0&\textbf{100}\\
  &DPMLE&0&\textbf{100}&0\\
  &DPMLE cov&0&\textbf{96}&4\\  \specialrule{.3em}{.2em}{0em}
  
\end{tabular}
\end{center}
\caption{\label{allresultsthree12000}
 Percentages of the $100$ simulation trials in 2-4 number of hidden states that were chosen by different criteria: AIC, BIC and DPMLE. "cov" refers to the models fitted with the linear covariate "time of day" in the transition probabilities, with $T=12,000$.}
\end{table}

\end{appendices}
\end{document}